\documentclass[aps, prd, floatfix, twocolumn, superscriptaddress, nofootinbib]{revtex4-1}

\usepackage{latexsym}
\usepackage{amsmath}
\usepackage{amssymb}
\usepackage{amsfonts}

\usepackage{slashed}
\usepackage{bm}

\usepackage[vcentermath]{youngtab}

\usepackage{color}
\definecolor{purple}{rgb}{0.5,0,0.5}
\definecolor{blue}{rgb}{0.0,0,0.9}
\definecolor{prdblue}{rgb}{0.133,0.118,0.498}

\usepackage{enumitem}
\usepackage{multirow}
\usepackage{subfigure}
\usepackage{textcomp}

\usepackage{supertabular} 
\usepackage{placeins}
\usepackage{epsfig}
\usepackage{graphicx}

\begin{document}

\title{Exploring the Efimov effect in the $D^*D^*D^*$ system}

\author{P. G. Ortega}
\email[]{pgortega@usal.es}
\affiliation{Departamento de Física Fundamental, Universidad de Salamanca, E-37008 Salamanca, Spain}
\affiliation{Instituto Universitario de F\'isica 
Fundamental y Matem\'aticas (IUFFyM), Universidad de Salamanca, E-37008 Salamanca, Spain}

\date{\today}

\begin{abstract}
The emergence of the Efimov effect in the $D^*D^*D^*$ system is explored under the assumption
that the heavy partner of the $T_{cc}^+$ exists as a $D^*D^*$ molecule with $(I)J^P=(0)1^+$.
The three-to-three relativistic scattering amplitude is obtained from the ladder amplitude formalism, built from
an energy-dependent contact two-body potential where the molecular component of the $T_{cc}^*$ state can be varied.
We find that $(I)J^P=(\tfrac{1}{2})0^-$ three-body bound states can be formed, with properties that suggest that the 
Efimov effect can be realised for reasonable values of the molecular probability and binding energy of the $T_{cc}^*$.
\end{abstract}



\maketitle



When two particles form a nearly resonant bound state due to short-range attractive forces, an effective long-range three-body emerges giving rise to an infinite number of three-body bound states with a discrete scale invariance. This remarkable phenomena, called \emph{Efimov effect}, was first described in the 1970's by V.~Efimov~\cite{Efimov:1970zz,2009NatPh...5..533E}. 
The Efimov effect has been mostly studied in atomic physics~\cite{Braaten:2006vd,Massignan:2008zza,Langmack:2017ubf,2017NatPh..13..731J,2009NatPh...5..227K}, due to its experimental observation in Cesium atoms in 2006~\cite{2006Natur.440..315K}. 
However, its relevance has also been explored in nuclear physics, e.g., in the $^{12}C$ three-$\alpha$ structure, the triton formation or the nuclear halo of $^{14}Be$, $^{22}C$ and $^{20}C$ nuclei~\cite{Bedaque:1999ve,Hammer:2008ra,Bedaque:1998kg,Mazumdar:1997zz,Amorim:1997mq,Mazumdar:2000dg}.

For sufficiently shallow two-body states the system becomes \emph{universal}, i.e., it is insensitive to the details of the short-range interaction and can be characterized by its $S$-wave scattering length $a_{\rm sc}$. In this limit, the three-body effective potential is proportional to $1/\rho^2$, where $\rho$ is the hyperspherical radius 
~\cite{Naidon:2016dpf,Braaten:2003he,Hammer:2010kp}. For three identical bosons of mass $m$ interacting via a short-range two-body potential the effective potential is attractive, so when $a_{\rm sc}\to\pm\infty$ an infinite family of three-body bound states appears with a scaling factor $\lambda=e^{\pi/|s_0|}\approx 22.7$. The ratio of the binding energies of the states satisfy $\Delta E^{(n)} \to Q^2\Delta E^{(n+1)}$, where $Q\to\lambda$ in the unitarity limit, i.e.,
\begin{align}
 \frac{\Delta E^{(n+1)}}{\Delta E^{(n)}} \to \lambda^{-2} \approx \frac{1}{515}
\end{align}
For large but finite scattering lengths the spectrum is not infinite, but few shallow Efimov states may emerge if some conditions are met~\cite{Naidon:2016dpf,Braaten:2004rn}.
In this case, the scaling law may also deviate from the universal value~\cite{PhysRevA.60.R9,Dawid:2023kxu}, so $Q\ne \lambda$.

The existence of three-body systems and the  possible appearance of the Efimov effect in hadronic physics has been also suggested in the recent literature~\cite{Braaten:2003he,Canham:2009zq,Wilbring:2017fwy,Valderrama:2018azi,MartinezTorres:2020hus}, specially since the discovery of the $X(3872)$ state~\cite{Belle:2003nnu}, a loosely-bound $D^{*\,0}\bar D^0$+h.c. molecule with quantum numbers $J^{PC}=1^{++}$. The properties of the $X(3872)$, unfortunately, rule out the existence of the Efimov effect~\cite{Braaten:2003he}.

However, the recent discovery of the $T_{cc}^+$~\cite{LHCb:2021vvq,LHCb:2021auc} can renew this interest. 
In 2021, the LHCb Collaboration discovered a new tetraquark-like state in the $D^0D^0\pi^+$ invariant mass spectrum~\cite{LHCb:2021vvq} with minimum quark content $cc\bar u\bar d$, named $T_{cc}^+$. The resonance is slightly below the $D^{*+}D^0$ threshold, with a binding energy estimated to be $\delta m_{\rm pole}=(360\pm40^{+4}_{-0})$~keV/c$^2$~\cite{LHCb:2021auc}. Its scattering length has a value of $a_{\rm sc}^{\rm LHCb}=-7.15(51)$ fm  and the experimental Weinberg factor~\footnote{This factor estimates the probability of finding a compact component in the wave function of a particle, with $Z=0$ for a pure molecular state and $Z=1$ for a pure compact state.} is  $Z<0.52(0.58)$ at $90(95)\%$ C.L. These properties of the $T_{cc}^+$ are compatible with a state with a sizable $DD^*$ molecular compound (for a review of the experimental and theoretical status of the $T_{cc}^+$ see, e.g., Ref.~\cite{Chen:2022asf} and references therein).

The announcement of the $T_{cc}^+$ has already stimulated the study of three-body states containing charmed mesons. For example, in Ref.~\cite{Wu:2021kbu} the authors explore the $DDD^*$ system, finding a bound state of few hundred keV with $(I)J^P=(\tfrac{1}{2})1^-$. Ref.~\cite{Pan:2022whr} analyze the existence of hadronic molecules composed of a $T_{cc}^{(*)}$ compact tetraquark and a $\bar D^{(*)}$ meson finding several candidates, while Refs~\cite{Luo:2021ggs,Bayar:2022bnc} study the $D^*D^*D^*$ and $DD^*D^*$ systems finding many potential three-body candidates in the $J^P=0^-$, $1^-$ and $2^-$ with isospins $\tfrac{1}{2}$ and $\tfrac{3}{2}$. However, no exploration of the Efimov effect in these systems has been suggested yet.

In this study we will explore the universality of the $D^*D^*D^*$ meson system in the $J^P=0^-$ sector with $I=\tfrac{1}{2}$, assuming that the isoscalar heavy partner of the $T_{cc}^+$, dubbed $T_{cc}^*$, exists close and below the $D^*D^*$ threshold, which is a prediction of Heavy-Quark Spin Symmetry (HQSS)~\cite{Neubert:1993mb}. HQSS implies that the heavy-meson interactions are insensitive to the spin of the heavy quark, so the interaction of the $D^*D^*$ system is identical to the $DD^*$ one for the $(I)J^P=(0)1^+$ sector (up to $1/M_Q$ corrections).
Actually, the $T_{cc}^*$ state has been predicted by many groups~\cite{Albaladejo:2021vln,Liu:2019stu,Dai:2021vgf,Du:2021zzh,Dong:2021bvy,Qiu:2023uno,Peng:2023lfw,Abreu:2022sra,Ortega:2022efc,Li:2012ss,Molina:2010tx,Jia:2022qwr,Whyte:2024ihh}, with a binding energy around few MeV.

The existence of the $T_{cc}^*$ with a small binding energy would favour the appearance of Efimov states in this sector.
If confirmed, it would be the first manifestation of the Efimov effect in hadronic physics, so it is worth exploring this system.
The choice of the $D^*D^*D^*$ system instead of the $DD^*D^*$ is that it allows us to work with identical bosons, simplifying the calculations. 
For three identical bosons, any two-body wave function must be symmetric. Then, the $J^P=0^-$ sector is selected because all the symmetric $D^*D^*$ pairs in S wave are in a relative $(I)J^P=(0)1^+$, whereas for $J^P=1^-$ and $2^-$ other isospin-spin $D^*D^*$ pairs are also allowed, such as the $(1)2^+$ and $(1)0^+$, adding repulsion to the three-body interaction~\cite{Bayar:2022bnc}.  Then, in this sector, all the allowed $D^*D^*$ pairs interact via an attractive potential, condition needed for the Efimov effect to emerge.


The starting point is the analysis of the $D^*D^*$ two-body system with a given binding energy ${\cal B}_2$, that is taken as a parameter. The two-body amplitude is obtained by solving the Bethe-Salpeter equation in the on-shell approximation~\cite{Nieves:1999bx},

\begin{align}\label{eq:BS}
 {\cal T}_2^{-1}(s) = {\cal V}^{-1}(s) - {\cal G}(s)\,,
\end{align}
where ${\cal V}(s)$ is the two-meson interaction and ${\cal G}$ is the relativistic two-meson loop function

\begin{align}
 {\cal G}(s) &= i\,\int \frac{d^4q}{(2\pi)^4} \frac{1}{q^2-m_1^2+i\varepsilon}\frac{1}{(P-q)^2-m_2^2+i\varepsilon}\,,
\end{align}
being $P$ the total initial four-momentum of the $D^*D^*$ meson system. This loop function is regularized via a sharp cutoff~\cite{Oller:1998hw}, i.e., assuming a maximum trimomentum module $|\vec q| \le \Lambda$. The value of the cutoff will be taken as $\Lambda=0.7$ GeV. However, we will analyze the sensitivity of the results by varying the cutoff between $\Lambda=0.5$ GeV and $1$ GeV.

We consider the same mass for $D^{*0}$ and $D^{*\pm}$, with $m=\tfrac{1}{2}(m_{D^{*0}}+m_{D^{*\pm}})=2008.55$ MeV. Then, for equal meson masses $m_1=m_2=m$, the loop function takes the form

\begin{align}\label{eq:loop2}
 {\cal G}(s)= \frac{1}{(4\pi)^2}&\left\{ \sigma\log\frac{\sigma\sqrt{1+\tfrac{m^2}{\Lambda^2}}+1}{\sigma\sqrt{1+\tfrac{m^2}{\Lambda^2}}-1} \right.\nonumber\\
 &\left.-2\log\left[\frac{\Lambda}{m}\left(1+\sqrt{1+\tfrac{m^2}{\Lambda^2}}\right)\right] \right\}\,,
\end{align}
with $\sigma=\sqrt{1-4m^2/s}$.
The prescription for the logarithmic function is such that Im(${\cal G}$) is given by

\begin{align}\label{eq:Gpres}
 {\rm Im}({\cal G}) &= -\frac{k}{8\pi\sqrt{s}}\Theta(s-4m^2)\,,
\end{align}
with $k$ the relativistic on-shell momentum for the $D^*D^*$ pair in its c.m. frame,
$ k = \tfrac{1}{2}\sqrt{s-4m^2}$.

The ${\cal V}$ potential for the $D^*D^*$ pair is taken as a $I=0$ $S$-wave interaction,
neglecting the $DD^*$-$D^*D^*$ coupled-channels effect and the finite width of the $D^*$.
These effects can be modelled as a source of width for the $T_{cc}^*$ and the $D^*D^*D^*$ trimers with little effect in the determination of the trimer masses~\cite{Dai:2021vgf,Bayar:2022bnc,Luo:2021ggs,Jia:2022qwr}. 

In order to evaluate the effect of the $T_{cc}^*$ inner composition, we follow Ref.~\cite{Montesinos:2023qbx} and consider the general energy-dependent contact potential

\begin{align}\label{eq:2bodypot}
 {\cal V}^{-1}(s) &= C_0 -C_1\frac{1-{\cal P}}{{\cal P}} (s-m_*^2),
\end{align}
being $C_0$ and $C_1$ constants and ${\cal P}$ the molecular probability in the $T_{cc}^*$ state, which ranges between $0$ and $1$.
These $C_0$ and $C_1$ parameters are fixed in order to impose the existence of a pole below the $D^*D^*$ threshold in the first Riemann sheet, with mass $m_*=2m-{\cal B}_2$. Their values are related to the loop function as $C_0\equiv {\cal G}(m_*^2)$ and $C_1\equiv{\cal G}^\prime(m_*^2)$, so there the only free parameters are ${\cal B}_2$ and ${\cal P}$. Depending on the molecular content in the $T_{cc}^*$ the potential ${\cal V}$ changes its energy dependence, being constant in the case of a pure $D^*D^*$ hadronic molecule (${\cal P}=1$) and behaving as ${\cal V}\sim\frac{1}{s-m_*^2}$ for a pure compact state (${\cal P}\to 0$).


The relativistic three-to-three scattering amplitude is built using the so-called \emph{ladder amplitude} formalism~\cite{Hansen:2015zga,Jackura:2020bsk,Dawid:2023jrj}. In this approach, the three-body amplitude is decomposed as~\footnote{The full $3\to 3$ scattering amplitude must be properly symmetrized, by summing over the nine possible spectator momenta (see, e.g., Ref~\cite{Jackura:2020bsk}), but in this work we refer to unsymmetrized amplitudes only.}

\begin{align}\label{eq1}
 {\cal M}_3(\vec p_i,\vec p_f) = {\cal D}(\vec p_i,\vec p_f)+{\cal M}_{df,3}(\vec p_i,\vec p_f)
\end{align}
where ${\cal D}$ is called the ladder amplitude, which contains the sum over all possible pairwise interactions connected through a sequence of one-particle exchanges, and ${\cal M}_{df,3}$ encodes all the contributions that arise when short-range three-particle interactions are present. The latter amplitude ${\cal M}_{df,3}$ depends on the ${\cal K}_{df,3}$ matrix, which represents the short-range three-body interactions~\cite{Dawid:2023kxu}. The $\vec p_i$ ($\vec p_f$) is the initial (final) momentum of one of the mesons, which is denoted as the \emph{spectator}. The other two remaining mesons associated with the given spectator are, then, called a pair or \emph{dimer}.
 If three-body short-range interactions are negligible, ${\cal K}_{df,3}={\cal M}_{df,3}=0$ and the ladder amplitude describes the full three-to-three scattering amplitude, ${\cal M}_3={\cal D}$.
In Refs.~\cite{MartinezTorres:2008gy,Khemchandani:2008rk} it was shown that there is a clean cancellation between the off-shell parts of the two-body $T$ matrix and the three-body forces in the framework of chiral Lagrangians. In this work we will make explicit use of this cancellation and neglect three-body forces, that's it, ${\cal K}_{df,3}=0$.
The equivalence of this method with alternative three-particle scattering formalism, e.g. the non-relativistic Faddeev equations, has been shown in Ref.~\cite{Jackura:2019bmu}.

In this work we assume that all the two-body subsystems are in a partial $S$ wave only, due to the proximity of the $T_{cc}^*$ state to the $D^*D^*$ threshold, which will suppress higher partial waves. At the same time, the dimer-spectator system is also assumed to be in $L=0$, which is the expected dominant partial wave~\cite{Bayar:2022bnc,Luo:2021ggs}. Indeed, in Ref.~\cite{Luo:2021ggs} the authors studied the $D^*D^*D^*$ system in all possible configurations with $L\le 2$ and found a small effect of the $S$-$D$ mixing compared to a $S$-wave only calculation.

\begin{figure}[t!]
\centering
 \includegraphics[width=.5\textwidth]{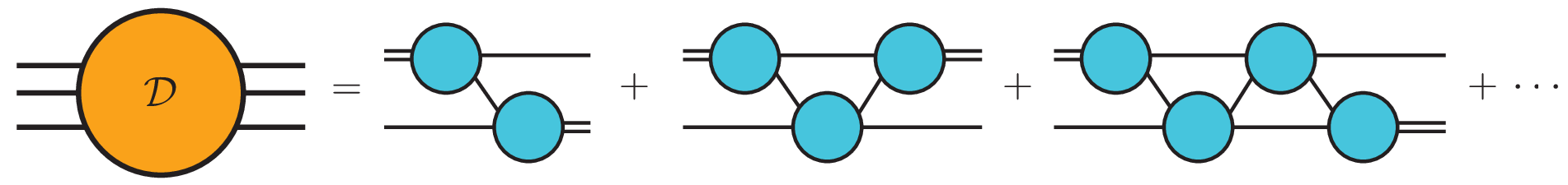}
 \caption{\label{fig1} Diagrammatic representation of the ladder amplitude ${\cal D}$ (Eq.~\eqref{eq2}), where the blue circles represent the two-body amplitude ${\cal M}_2$ and the diagonal black lines connecting them are the one-particle exchange function $G$.}
\end{figure}

The ${\cal D}$ ladder amplitude is defined by the integral equation

\begin{align}\label{eq2}
 {\cal D}(\vec p_i,\vec p_f) =& -{\cal M}_2(p_i)G(\vec p_i,\vec p_f){\cal M}_2(p_f) \nonumber \\
		&	-{\cal M}_2(p_i)\int \frac{d^3\vec q}{(2\pi)^32\omega(q)}G(\vec p_i,\vec q){\cal D}(\vec q,\vec p_f)\,,
\end{align}
which is diagrammatically shown in Fig.~\ref{fig1}.

Here, $G$ is the long-range interaction between the dimer and the spectator, mediated by a particle exchange, and ${\cal M}_2$ is the relativistic $2\to 2$ scattering amplitude describing the meson-meson interaction in the dimer. The dimer energy is fixed by the momentum of the spectator, $s_2(p)=(\sqrt{s}-\omega(p))^2-p^2$, with $\omega(p)=\sqrt{m^2+p^2}$ and $p=|\vec p|$. The ${\cal M}_2$ amplitude is proportional to the two-body $T$ matrix in Eq.~\eqref{eq:BS} as ${\cal M}_2=-2\,{\cal T}_2$.

For all pairs in $S$ wave, the one-particle exchange propagator (OPE) $G$ can be written as

\begin{align}\label{eq:OPE}
 G(p_i,p_f) &\equiv -\frac{H(p_i,p_f)}{4p_ip_f}\log\left(\frac{z(p_i,p_f)-2p_ip_f+i\epsilon}{z(p_i,p_f)+2p_ip_f+i\epsilon}\right)\,,
\end{align}
where $z(p_i,p_f)=(\sqrt{s}-\omega(p_i)-\omega(p_f))^2-p_i^2-p_f^2-m^2$ and $H(p_i,p_f)$ is a cutoff function to ensure a finite integral in Eq.~\eqref{eq2}. For this cutoff function we use a sharp cutoff, which is one in the integration domain and zero elsewhere.

It is worth noticing that we are studying a system of three vector mesons with isospin $\tfrac{1}{2}$, different from the spinless case studied in Refs.~\cite{Hansen:2015zga,Jackura:2020bsk,Dawid:2023jrj}. The recoupling coefficient with particles with spin is more delicate than the spinless case~\cite{PhysRevC.32.502}, but for dimer-spectator systems in $S$-wave,
the OPE of a $D^*$ between two $(0)1^+$ dimers can be reduced to the Eq.~\eqref{eq:OPE} multiplied by the
spin-isospin recoupling~\cite{Jackura:2023qtp}

\begin{align}
 G(p_i,p_f) \longrightarrow \langle I_2,I|I_2',I\rangle \langle S_2,S|S_2',S\rangle\, G(p_i,p_f)
\end{align}
where $S$ ($I$) is the total three-body spin (isospin) and $S_2^{(\prime)}$ ($I_2^{(\prime)}$) is the spin (isospin) of the initial and final dimer.

%
For the $D^*D^*D^*$ in $J^P=0^-$, this implies $G(p_i,p_f) \to \tfrac{1}{2} G(p_i,p_f)$.
In addition, an extra factor $2$ must be added in the ${\cal M}_2$ to account for the different isospin projections of the initial/final $D^*D^*$ dimers, that's it, ${\cal M}_2(p) \to 2{\cal M}_2(p)$.

For three identical bosons, it is more convenient to work with the amputated amplitude $d$, 

\begin{align}
 {\cal D}(\vec p_i,\vec p_f) \equiv {\cal M}_2(p_i)d(\vec p_i,\vec p_f){\cal M}_2(p_f)\,,
\end{align}
which eliminates the singularities of the dimer ${\cal M}_2$ amplitude.
With this definition, the $S$-wave amputated amplitude is given by

\begin{align}\label{eq3}
 d(p_i,p_f) =& -G(p_i, p_f) \nonumber\\&
			-\int \frac{dq\,q^2}{(2\pi)^32\omega(q)}G(p_i,q){\cal M}_2(q) d(q,p_f)\,,
\end{align}
which will be numerically solved following Ref.~\cite{Dawid:2023jrj}.


As mentioned above, the main purpose of this work is to evaluate  the existence of Efimov states in the $D^*D^*D^*$ in the $(I)J^P=(\tfrac{1}{2})0^-$ sector, given the existence of a bound $(I)J^P=(0)1^+$ $D^*D^*$ state, $T_{cc}^*$, which has been predicted as the HQSS partner of the $T_{cc}^+$ state recently discovered.
The advantage of using this system is that we can work with identical bosons, which favours the generation of Efimov states when the two-body pairs have attractive nearly-resonant interactions. However, the main uncertainty are the properties of this hypothetical $T_{cc}^*$, i.e., how close we are to the resonant limit.
For this reason we evaluate the possible $D^*D^*D^*$ trimer states assuming a selection of $T_{cc}^*$ binding energies, ${\cal B}_2=\{0.01,0.5,1.0,5.0\}$ MeV. Then, we will study their properties as a function of the $T_{cc}^*$ molecular content ${\cal P}$.

\begin{table}[t!]
\centering
\caption{\label{tab:ERE} Properties of the trimer states in the effective range expansion approach for the 2-body amplitude. \emph{ 1$^{st}$ column:} Two-body binding energy, in MeV; \emph{2$^{nd}$ column: } Two-body scattering length, in fm; \emph{3$^{rd}$ to 5$^{th}$ columns: } Binding energies of the i$^{th}$ trimer state, ${\cal B}_3^{(i)}=3m-E^{(i)}$, with $E^{(i)}$ the 3-body mass of the i$^{th}$ trimer, in MeV; \emph{6$^{th}$ column: } Ratio of the second to first trimer binding energies. The dagger ($\dagger$) indicates a virtual state (pole in the second Riemann sheet). }
\begin{ruledtabular}
\begin{tabular}{cccccc}
${\cal B}_2$ 	&	$a_{\rm sc}$ 	&	${\cal B}_3^{(1)}$	&	${\cal B}_3^{(2)}$	& ${\cal B}_3^{(3)}$ &	${\cal B}_3^{(2)}/{\cal B}_3^{(1)}$\\
\hline
$0.01$ & $44.03$ & $54.592$ & $0.185$ & $0.011$  & $0.003$\\
$0.5$ & $6.23$ & $64.158$ & $0.980$ &  $0.620$$^\dagger$ & $0.015$\\
$1.0$ & $4.40$ & $69.099$ & $1.557$ & $-$ & $0.023$ \\
$5.0$ & $1.97$ & $91.365$ & $5.521$ & $-$ & $0.060$ \\
\end{tabular}
\end{ruledtabular}
\end{table}

\begin{figure}[t!]
 \includegraphics[width=.48\textwidth]{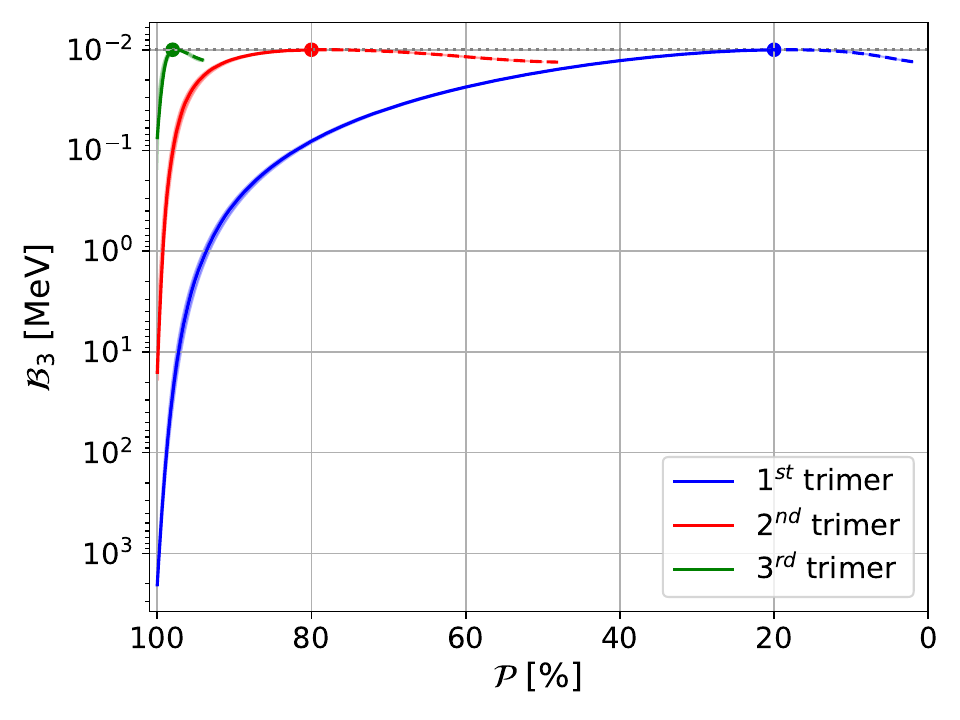}
 \includegraphics[width=.48\textwidth]{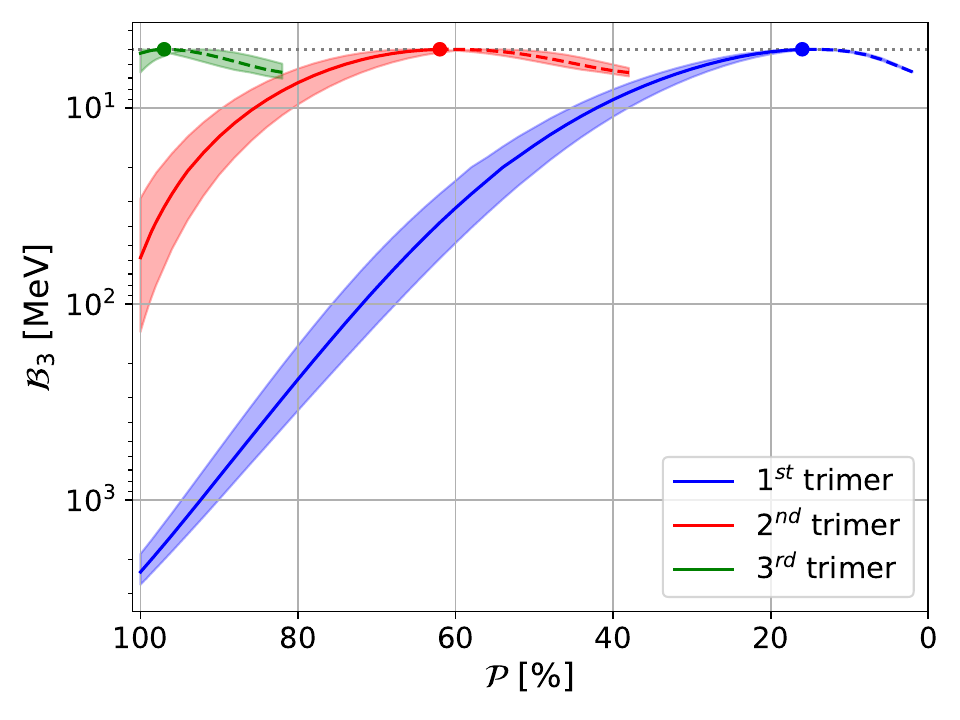}
 \caption{\label{fig:trimers} Binding energies of the first $D^*D^*D^*$ trimers (${\cal B}_3=3m-E_3$) for ${\cal B}_2=0.01$ MeV (upper panel) and ${\cal B}_2=5$ MeV (lower panel) as a function of the $T_{cc}^*$ composition, ranging from a purely two-body molecular state (${\cal P}=100\%$) to a purely compact $T_{cc}^*$ state (${\cal P}=0\%$). The central lines show the results for $\Lambda=0.7$ GeV cutoff in the two-body amplitude. The color error bands indicate the results for the cutoff range $\Lambda=[0.5,1]$ GeV. Solid lines represent bound states, whereas dashed lines represent virtual states. The dot marks the value of ${\cal P}$ where the pole changes the Riemann sheet. The dotted horizontal gray line shows the two-body binding energy ${\cal B}_2$, which acts as the threshold for the trimer states. }
\end{figure}

A first simple study that can provide useful insights into the problem is the analysis in the leading-order effective range expansion (ERE), that's it,

\begin{align}
{\cal T}_2(p) =& \frac{8\pi\sqrt{s_2(p)}}{iq_2(p)+1/a_{\rm sc}}\,,
\end{align}
with $q_2(p)=\sqrt{s_2(p)/4-m^2}$ the relative momentum of the particles in the dimer.
In this case, the phenomenology of the three-body states only depend on the $D^*D^*$ scattering length $a_{\rm sc} = 2\hbar c/\sqrt{4m^2-m_*^2}$.
The results of this approach are given in Table~\ref{tab:ERE}. For the four ${\cal B}_2$ under evaluation we find, at least, two bound-state trimers. For ${\cal B}_2=0.01$ MeV and $0.5$ MeV a third bound and virtual trimer state is also found, respectively. Of course, the smaller the binding energy, the closer we are to the resonant limit. These results agree with the Efimov states analyzed in Ref.~\cite{Dawid:2023kxu} for the general case of three identical bosons, both their energies and the ratios of subsequent binding energies. This suggests that, indeed, the Efimov effect can be present in the $D^*D^*D^*$ system.

The results are, though, more interesting when the full two-body potential introduced in Eq.~\eqref{eq:2bodypot} is used. In this case, the first three Efimov trimers are also found, but their masses depend on the molecular probability of the $T_{cc}^*$. The masses of the $D^*D^*D^*$ trimers for ${\cal B}_2=0.01$ MeV and $5$ MeV are shown in the upper and lower panels of Fig.~\ref{fig:trimers}, respectively. Generally, the larger ${\cal P}$ and ${\cal B}_2$ in the $T_{cc}^*$, the deeper the binding energy of the trimers. For ${\cal P}>20\%$ the first trimer emerges as a bound state, while the second emerges between $60\%$ and $80\%$, depending on ${\cal B}_2$.
Contrary to the ERE results, three Efimov states are found for all binding energies, but ${\cal P}$ values above $96\%$ are needed in order to have three bound states, so it is unlikely that the third trimer will exist unless the $T_{cc}^*$ is a pure molecule. 

It should be recalled that the three body binding energies are reduced with ${\cal P}$ because the coupling of the $T_{cc}^*$ with the remaining $D^*$ drops to zero as ${\cal P}\to 0$, as it does the scattering length $a_{\rm sc}$. This formalism does not include any interactions between a given compact component of the $T_{cc}^*$ and the charmed meson, which could add further sources of attraction or repulsion as explored in Ref.~\cite{Pan:2022whr}. Needless to say, if the $T_{cc}^*$ is a compact tetraquark the $T_{cc}^*D^*$ system becomes a two-body problem and no Efimov state would manifest.

\begin{figure}[t!]
 \includegraphics[width=.48\textwidth]{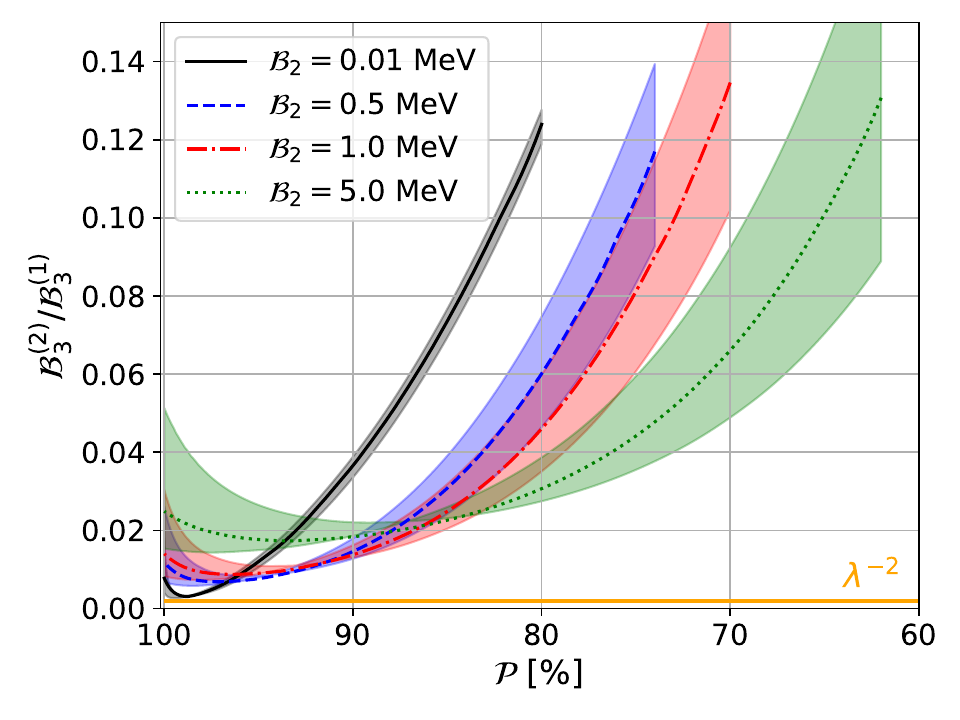}
 \caption{\label{fig:ratios} Ratio of the second to first trimer binding energies for different ${\cal B}_2$ as a function of the $T_{cc}^*$ composition, where ${\cal P}=100\%$ denotes a pure two-body $T_{cc}^*$ molecule and ${\cal P}=0\%$ a pure compact state. The horizontal orange line represents the Efimov scaling factor  $\lambda^{-2}\sim 1/515$ at the unitary limit $a_{\rm sc}\to\infty$. A cutoff of $\Lambda=0.7$ GeV has been used in Eq.~\eqref{eq:loop2} for the central line, while the color error bands indicate the results for the cutoff range $\Lambda=[0.5,1]$ GeV.}
\end{figure}

The ratios of the binding energies of the first and second trimer are shown in Fig.~\ref{fig:ratios}. The results indicate that the ratio increases as the $T_{cc}^*$ becomes more compact, with a small valley that gets shallower as ${\cal B}_2$ increases. The bottom of this valley approaches the Efimov scaling law $\lambda^{-2}\sim 1/515 \sim 0.0019$ the smaller ${\cal B}_2$ and it is expected to reach it for ${\cal B}_2\to 0$.  This deviation from the $\lambda^{-2}$ universal value appears because we are not exactly in the unitarity limit~\cite{Dawid:2023kxu}. Furthermore, the addition of possible compact components in the $T_{cc}^*$ wave function that mix with the $D^*D^*$ pairs further modifies this scaling.

\begin{figure}[t!]
 \includegraphics[width=.48\textwidth]{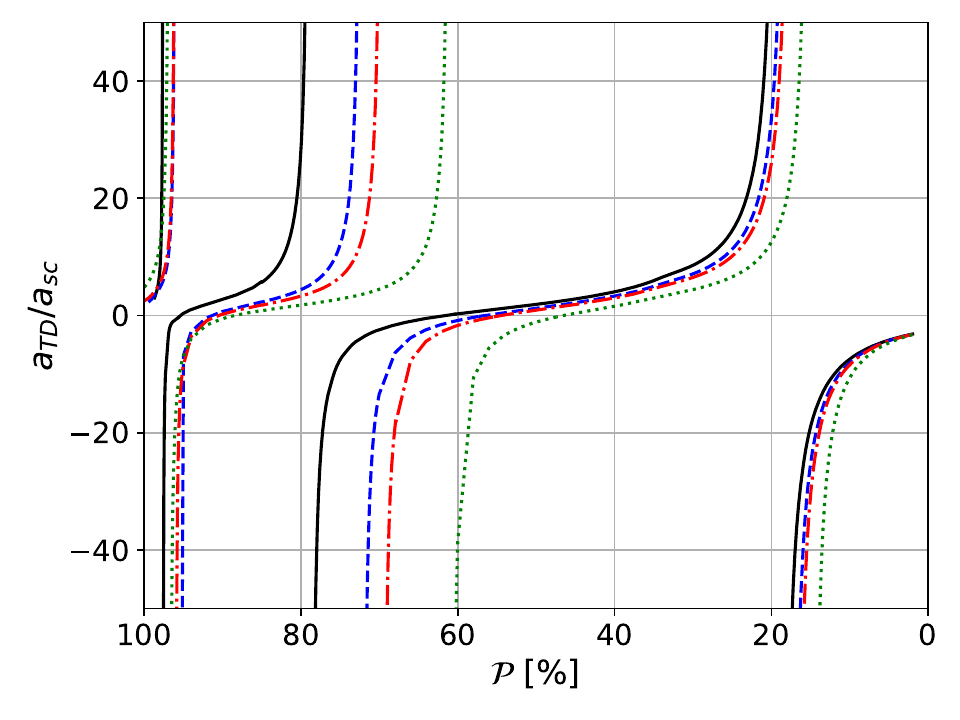}
 \caption{\label{fig:aTD} $T_{cc}^*D^*$ scattering length normalized over the $D^*D^*$ scattering length as a function of the $T_{cc}^*$ composition for different binding energies ${\cal B}_2$, using $\Lambda=0.7$ GeV. Same legend as in Fig.~\ref{fig:ratios}.}
\end{figure}

As we have seen, the upper limit for the trimer states is the $T_{cc}^*D^*$ threshold.
Actually, this channel can be a potential detection mechanism. Indeed, the existence of the $T_{cc}^*$ allows us to evaluate the scattering length of the $T_{cc}^*$ and $D^*$, which will be called $a_{\rm TD}$.
This can be calculated by using the dimer-spectator scattering amplitude ${\cal M}_{\rm TD}$, which encodes the information of the $T_{cc}^*D^*\to T_{cc}^*D^*$ reaction and is obtained by expanding the three-body amplitude ${\cal M}_3$ in the vicinity of the two-body bound state $m_*$~\cite{Jackura:2020bsk},

\begin{align}
 {\cal M}_{\rm TD}(s) = g^2 \lim_{s_2(p_i),s_2(p_f)\to m_*^2}d(p_i,p_f)
\end{align}
where $g^2$ is the residue of the 2-body scattering amplitude ${\cal M}_2$ around the $T_{cc}^*$ mass.
Then, the $T_{cc}^*D^*$ scattering length can be calculated as,

\begin{align}
-\frac{1}{a_{\rm TD}} =\lim_{s\to m_{\rm TD}^2} 8\pi \sqrt{s} Re({\cal M}_{\rm TD}^{-1}(s))
\end{align}
with $m_{\rm TD}=3m-{\cal B}_2$ the mass of the $T_{cc}^*D^*$ threshold. Results for $a_{\rm TD}$ for different values of ${\cal B}_2$ and ${\cal P}$ are shown in Fig.~\ref{fig:aTD}.
The crossing of the $T_{cc}^*D^*$ threshold by the $D^*D^*D^*$ states is shown as an infinite in $a_{\rm TD}$, similarly to a two-body state.
Then, this parameter can give us information about the $T_{cc}^*D^*$ scattering and the closeness of a pole. 


In this work we have analyzed the $D^*D^*D^*$ system in the $(I)J^P=(\tfrac{1}{2})0^-$ sector.
The results indicate that the Efimov effect can indeed emerge in this system. That's it, we find a spectrum of trimer states bound due to long-range interactions as a consequence of a nearly-resonant two-body system, explored using a general energy-dependent two body potential that models a mixed $T_{cc}^*$ state composed of compact and molecular $D^*D^*$ structures. At least one trimer can be formed, as predicted by some studies~\cite{Luo:2021ggs,Bayar:2022bnc}. The emergence of a second and third trimer depends on the molecular percentage of the $T_{cc}^*$ resonance and its binding energy. The third trimer is unlikely to be bound, but the second one can be formed for reasonable ${\cal P}$ and ${\cal B}_2$. Thus, this system deserves more experimental and theoretical studies to clarify this phenomena. Of course, the first step would be the experimental detection of the $T_{cc}^*$ state.

We want to remark that we don't discard this effect in the $DD^*D^*$ system. According to HQSS, the $DD^*$ and $D^*D^*$ systems have the same potential in the $(0)1^+$ sector. Then, the scaling law of the Efimov effect in the case of two identical bosons $M$ plus one particle $m$ resonantly interacting with each other is also $\lambda\approx 22.7$. Actually, this ratio decreases with the mass ratio $M/m$, but for the $DD^*D^*$ the factor is $m_{D^*}/m_D\approx 1.08$ and the effect will be small (See Fig.~12 of Ref.~\cite{Naidon:2016dpf}). However, a more detailed calculation would be needed in order to fully clarify the existence of trimers in the $T_{cc}D^*$-$T_{cc}^*D$ systems.

The realisation of the Efimov effect with charmed mesons would be an exceptional discovery and a step forward in our understanding of multimeson states, and this system is a promising place to investigate it. In fact, an interesting framework could be the analysis of the $D^*D^*D^*$ system in nuclear medium, as some studies show significant modifications of masses and widths of the $T_{cc}^{(*)}$ state~\cite{Montesinos:2023qbx}, so it is possible that the resonant limit can be modulated for specific nuclear densities.


\begin{acknowledgments}
The author wants to thank Eulogio Oset, Andrew~W.~Jackura and Ra\'ul~A.~Brice\~no for fruitful discussions.
This work has been partially funded by
EU Horizon 2020 research and innovation program, STRONG-2020 project, under grant agreement no. 824093; and
Ministerio Espa\~nol de Ciencia e Innovaci\'on under grant Nos. PID2019-105439GB-C22 and PID2022-141910NB-I00.
\end{acknowledgments}


\bibliographystyle{apsrev4-1} 
\bibliography{Biblio}

\begin{thebibliography}{57}%
\makeatletter
\providecommand \@ifxundefined [1]{%
 \@ifx{#1\undefined}
}%
\providecommand \@ifnum [1]{%
 \ifnum #1\expandafter \@firstoftwo
 \else \expandafter \@secondoftwo
 \fi
}%
\providecommand \@ifx [1]{%
 \ifx #1\expandafter \@firstoftwo
 \else \expandafter \@secondoftwo
 \fi
}%
\providecommand \natexlab [1]{#1}%
\providecommand \enquote  [1]{``#1''}%
\providecommand \bibnamefont  [1]{#1}%
\providecommand \bibfnamefont [1]{#1}%
\providecommand \citenamefont [1]{#1}%
\providecommand \href@noop [0]{\@secondoftwo}%
\providecommand \href [0]{\begingroup \@sanitize@url \@href}%
\providecommand \@href[1]{\@@startlink{#1}\@@href}%
\providecommand \@@href[1]{\endgroup#1\@@endlink}%
\providecommand \@sanitize@url [0]{\catcode `\\12\catcode `\$12\catcode
  `\&12\catcode `\#12\catcode `\^12\catcode `\_12\catcode `\%12\relax}%
\providecommand \@@startlink[1]{}%
\providecommand \@@endlink[0]{}%
\providecommand \url  [0]{\begingroup\@sanitize@url \@url }%
\providecommand \@url [1]{\endgroup\@href {#1}{\urlprefix }}%
\providecommand \urlprefix  [0]{URL }%
\providecommand \Eprint [0]{\href }%
\providecommand \doibase [0]{http://dx.doi.org/}%
\providecommand \selectlanguage [0]{\@gobble}%
\providecommand \bibinfo  [0]{\@secondoftwo}%
\providecommand \bibfield  [0]{\@secondoftwo}%
\providecommand \translation [1]{[#1]}%
\providecommand \BibitemOpen [0]{}%
\providecommand \bibitemStop [0]{}%
\providecommand \bibitemNoStop [0]{.\EOS\space}%
\providecommand \EOS [0]{\spacefactor3000\relax}%
\providecommand \BibitemShut  [1]{\csname bibitem#1\endcsname}%
\let\auto@bib@innerbib\@empty
\bibitem [{\citenamefont {Efimov}(1970)}]{Efimov:1970zz}%
  \BibitemOpen
  \bibfield  {author} {\bibinfo {author} {\bibfnamefont {V.}~\bibnamefont
  {Efimov}},\ }\href {\doibase 10.1016/0370-2693(70)90349-7} {\bibfield
  {journal} {\bibinfo  {journal} {Phys. Lett. B}\ }\textbf {\bibinfo {volume}
  {33}},\ \bibinfo {pages} {563} (\bibinfo {year} {1970})}\BibitemShut
  {NoStop}%
\bibitem [{\citenamefont {{Efimov}}(2009)}]{2009NatPh...5..533E}%
  \BibitemOpen
  \bibfield  {author} {\bibinfo {author} {\bibfnamefont {V.}~\bibnamefont
  {{Efimov}}},\ }\href {\doibase 10.1038/nphys1355} {\bibfield  {journal}
  {\bibinfo  {journal} {Nature Physics}\ }\textbf {\bibinfo {volume} {5}},\
  \bibinfo {pages} {533} (\bibinfo {year} {2009})}\BibitemShut {NoStop}%
\bibitem [{\citenamefont {Braaten}\ and\ \citenamefont
  {Hammer}(2007)}]{Braaten:2006vd}%
  \BibitemOpen
  \bibfield  {author} {\bibinfo {author} {\bibfnamefont {E.}~\bibnamefont
  {Braaten}}\ and\ \bibinfo {author} {\bibfnamefont {H.~W.}\ \bibnamefont
  {Hammer}},\ }\href {\doibase 10.1016/j.aop.2006.10.011} {\bibfield  {journal}
  {\bibinfo  {journal} {Annals Phys.}\ }\textbf {\bibinfo {volume} {322}},\
  \bibinfo {pages} {120} (\bibinfo {year} {2007})},\ \Eprint
  {http://arxiv.org/abs/cond-mat/0612123} {arXiv:cond-mat/0612123} \BibitemShut
  {NoStop}%
\bibitem [{\citenamefont {Massignan}\ and\ \citenamefont
  {Stoof}(2008)}]{Massignan:2008zza}%
  \BibitemOpen
  \bibfield  {author} {\bibinfo {author} {\bibfnamefont {P.}~\bibnamefont
  {Massignan}}\ and\ \bibinfo {author} {\bibfnamefont {H.~T.~C.}\ \bibnamefont
  {Stoof}},\ }\href {\doibase 10.1103/PhysRevA.78.030701} {\bibfield  {journal}
  {\bibinfo  {journal} {Phys. Rev. A}\ }\textbf {\bibinfo {volume} {78}},\
  \bibinfo {pages} {030701} (\bibinfo {year} {2008})},\ \Eprint
  {http://arxiv.org/abs/cond-mat/0702462} {arXiv:cond-mat/0702462} \BibitemShut
  {NoStop}%
\bibitem [{\citenamefont {Langmack}\ \emph {et~al.}(2018)\citenamefont
  {Langmack}, \citenamefont {Schmidt},\ and\ \citenamefont
  {Zwerger}}]{Langmack:2017ubf}%
  \BibitemOpen
  \bibfield  {author} {\bibinfo {author} {\bibfnamefont {C.}~\bibnamefont
  {Langmack}}, \bibinfo {author} {\bibfnamefont {R.}~\bibnamefont {Schmidt}}, \
  and\ \bibinfo {author} {\bibfnamefont {W.}~\bibnamefont {Zwerger}},\ }\href
  {\doibase 10.1103/PhysRevA.97.033623} {\bibfield  {journal} {\bibinfo
  {journal} {Phys. Rev. A}\ }\textbf {\bibinfo {volume} {97}},\ \bibinfo
  {pages} {033623} (\bibinfo {year} {2018})},\ \Eprint
  {http://arxiv.org/abs/1709.00749} {arXiv:1709.00749 [cond-mat.quant-gas]}
  \BibitemShut {NoStop}%
\bibitem [{\citenamefont {{Johansen}}\ \emph {et~al.}(2017)\citenamefont
  {{Johansen}}, \citenamefont {{Desalvo}}, \citenamefont {{Patel}},\ and\
  \citenamefont {{Chin}}}]{2017NatPh..13..731J}%
  \BibitemOpen
  \bibfield  {author} {\bibinfo {author} {\bibfnamefont {J.}~\bibnamefont
  {{Johansen}}}, \bibinfo {author} {\bibfnamefont {B.~J.}\ \bibnamefont
  {{Desalvo}}}, \bibinfo {author} {\bibfnamefont {K.}~\bibnamefont {{Patel}}},
  \ and\ \bibinfo {author} {\bibfnamefont {C.}~\bibnamefont {{Chin}}},\ }\href
  {\doibase 10.1038/nphys4130} {\bibfield  {journal} {\bibinfo  {journal}
  {Nature Physics}\ }\textbf {\bibinfo {volume} {13}},\ \bibinfo {pages} {731}
  (\bibinfo {year} {2017})},\ \Eprint {http://arxiv.org/abs/1612.05169}
  {arXiv:1612.05169 [cond-mat.quant-gas]} \BibitemShut {NoStop}%
\bibitem [{\citenamefont {{Knoop}}\ \emph {et~al.}(2009)\citenamefont
  {{Knoop}}, \citenamefont {{Ferlaino}}, \citenamefont {{Mark}}, \citenamefont
  {{Berninger}}, \citenamefont {{Sch{\"o}bel}}, \citenamefont {{N{\"a}gerl}},\
  and\ \citenamefont {{Grimm}}}]{2009NatPh...5..227K}%
  \BibitemOpen
  \bibfield  {author} {\bibinfo {author} {\bibfnamefont {S.}~\bibnamefont
  {{Knoop}}}, \bibinfo {author} {\bibfnamefont {F.}~\bibnamefont {{Ferlaino}}},
  \bibinfo {author} {\bibfnamefont {M.}~\bibnamefont {{Mark}}}, \bibinfo
  {author} {\bibfnamefont {M.}~\bibnamefont {{Berninger}}}, \bibinfo {author}
  {\bibfnamefont {H.}~\bibnamefont {{Sch{\"o}bel}}}, \bibinfo {author}
  {\bibfnamefont {H.~C.}\ \bibnamefont {{N{\"a}gerl}}}, \ and\ \bibinfo
  {author} {\bibfnamefont {R.}~\bibnamefont {{Grimm}}},\ }\href {\doibase
  10.1038/nphys1203} {\bibfield  {journal} {\bibinfo  {journal} {Nature
  Physics}\ }\textbf {\bibinfo {volume} {5}},\ \bibinfo {pages} {227} (\bibinfo
  {year} {2009})},\ \Eprint {http://arxiv.org/abs/0807.3306} {arXiv:0807.3306
  [cond-mat.other]} \BibitemShut {NoStop}%
\bibitem [{\citenamefont {{Kraemer}}\ \emph {et~al.}(2006)\citenamefont
  {{Kraemer}}, \citenamefont {{Mark}}, \citenamefont {{Waldburger}},
  \citenamefont {{Danzl}}, \citenamefont {{Chin}}, \citenamefont {{Engeser}},
  \citenamefont {{Lange}}, \citenamefont {{Pilch}}, \citenamefont {{Jaakkola}},
  \citenamefont {{N{\"a}gerl}},\ and\ \citenamefont
  {{Grimm}}}]{2006Natur.440..315K}%
  \BibitemOpen
  \bibfield  {author} {\bibinfo {author} {\bibfnamefont {T.}~\bibnamefont
  {{Kraemer}}}, \bibinfo {author} {\bibfnamefont {M.}~\bibnamefont {{Mark}}},
  \bibinfo {author} {\bibfnamefont {P.}~\bibnamefont {{Waldburger}}}, \bibinfo
  {author} {\bibfnamefont {J.~G.}\ \bibnamefont {{Danzl}}}, \bibinfo {author}
  {\bibfnamefont {C.}~\bibnamefont {{Chin}}}, \bibinfo {author} {\bibfnamefont
  {B.}~\bibnamefont {{Engeser}}}, \bibinfo {author} {\bibfnamefont {A.~D.}\
  \bibnamefont {{Lange}}}, \bibinfo {author} {\bibfnamefont {K.}~\bibnamefont
  {{Pilch}}}, \bibinfo {author} {\bibfnamefont {A.}~\bibnamefont {{Jaakkola}}},
  \bibinfo {author} {\bibfnamefont {H.~C.}\ \bibnamefont {{N{\"a}gerl}}}, \
  and\ \bibinfo {author} {\bibfnamefont {R.}~\bibnamefont {{Grimm}}},\ }\href
  {\doibase 10.1038/nature04626} {\bibfield  {journal} {\bibinfo  {journal}
  {\nat}\ }\textbf {\bibinfo {volume} {440}},\ \bibinfo {pages} {315} (\bibinfo
  {year} {2006})},\ \Eprint {http://arxiv.org/abs/cond-mat/0512394}
  {arXiv:cond-mat/0512394 [cond-mat.other]} \BibitemShut {NoStop}%
\bibitem [{\citenamefont {Bedaque}\ \emph {et~al.}(2000)\citenamefont
  {Bedaque}, \citenamefont {Hammer},\ and\ \citenamefont {van
  Kolck}}]{Bedaque:1999ve}%
  \BibitemOpen
  \bibfield  {author} {\bibinfo {author} {\bibfnamefont {P.~F.}\ \bibnamefont
  {Bedaque}}, \bibinfo {author} {\bibfnamefont {H.~W.}\ \bibnamefont {Hammer}},
  \ and\ \bibinfo {author} {\bibfnamefont {U.}~\bibnamefont {van Kolck}},\
  }\href {\doibase 10.1016/S0375-9474(00)00205-0} {\bibfield  {journal}
  {\bibinfo  {journal} {Nucl. Phys. A}\ }\textbf {\bibinfo {volume} {676}},\
  \bibinfo {pages} {357} (\bibinfo {year} {2000})},\ \Eprint
  {http://arxiv.org/abs/nucl-th/9906032} {arXiv:nucl-th/9906032} \BibitemShut
  {NoStop}%
\bibitem [{\citenamefont {Hammer}\ and\ \citenamefont
  {Higa}(2008)}]{Hammer:2008ra}%
  \BibitemOpen
  \bibfield  {author} {\bibinfo {author} {\bibfnamefont {H.~W.}\ \bibnamefont
  {Hammer}}\ and\ \bibinfo {author} {\bibfnamefont {R.}~\bibnamefont {Higa}},\
  }\href {\doibase 10.1140/epja/i2008-10617-3} {\bibfield  {journal} {\bibinfo
  {journal} {Eur. Phys. J. A}\ }\textbf {\bibinfo {volume} {37}},\ \bibinfo
  {pages} {193} (\bibinfo {year} {2008})},\ \Eprint
  {http://arxiv.org/abs/0804.4643} {arXiv:0804.4643 [nucl-th]} \BibitemShut
  {NoStop}%
\bibitem [{\citenamefont {Bedaque}\ \emph {et~al.}(1999)\citenamefont
  {Bedaque}, \citenamefont {Hammer},\ and\ \citenamefont {van
  Kolck}}]{Bedaque:1998kg}%
  \BibitemOpen
  \bibfield  {author} {\bibinfo {author} {\bibfnamefont {P.~F.}\ \bibnamefont
  {Bedaque}}, \bibinfo {author} {\bibfnamefont {H.~W.}\ \bibnamefont {Hammer}},
  \ and\ \bibinfo {author} {\bibfnamefont {U.}~\bibnamefont {van Kolck}},\
  }\href {\doibase 10.1103/PhysRevLett.82.463} {\bibfield  {journal} {\bibinfo
  {journal} {Phys. Rev. Lett.}\ }\textbf {\bibinfo {volume} {82}},\ \bibinfo
  {pages} {463} (\bibinfo {year} {1999})},\ \Eprint
  {http://arxiv.org/abs/nucl-th/9809025} {arXiv:nucl-th/9809025} \BibitemShut
  {NoStop}%
\bibitem [{\citenamefont {Mazumdar}\ and\ \citenamefont
  {Bhasin}(1997)}]{Mazumdar:1997zz}%
  \BibitemOpen
  \bibfield  {author} {\bibinfo {author} {\bibfnamefont {I.}~\bibnamefont
  {Mazumdar}}\ and\ \bibinfo {author} {\bibfnamefont {V.~S.}\ \bibnamefont
  {Bhasin}},\ }\href {\doibase 10.1103/PhysRevC.56.R5} {\bibfield  {journal}
  {\bibinfo  {journal} {Phys. Rev. C}\ }\textbf {\bibinfo {volume} {56}},\
  \bibinfo {pages} {R5} (\bibinfo {year} {1997})}\BibitemShut {NoStop}%
\bibitem [{\citenamefont {Amorim}\ \emph {et~al.}(1997)\citenamefont {Amorim},
  \citenamefont {Frederico},\ and\ \citenamefont {Tomio}}]{Amorim:1997mq}%
  \BibitemOpen
  \bibfield  {author} {\bibinfo {author} {\bibfnamefont {A.~E.~A.}\
  \bibnamefont {Amorim}}, \bibinfo {author} {\bibfnamefont {T.}~\bibnamefont
  {Frederico}}, \ and\ \bibinfo {author} {\bibfnamefont {L.}~\bibnamefont
  {Tomio}},\ }\href@noop {} {\  (\bibinfo {year} {1997})},\ \Eprint
  {http://arxiv.org/abs/nucl-th/9708023} {arXiv:nucl-th/9708023} \BibitemShut
  {NoStop}%
\bibitem [{\citenamefont {Mazumdar}\ \emph {et~al.}(2000)\citenamefont
  {Mazumdar}, \citenamefont {Arora},\ and\ \citenamefont
  {Bhasin}}]{Mazumdar:2000dg}%
  \BibitemOpen
  \bibfield  {author} {\bibinfo {author} {\bibfnamefont {I.}~\bibnamefont
  {Mazumdar}}, \bibinfo {author} {\bibfnamefont {V.}~\bibnamefont {Arora}}, \
  and\ \bibinfo {author} {\bibfnamefont {V.~S.}\ \bibnamefont {Bhasin}},\
  }\href {\doibase 10.1103/PhysRevC.61.051303} {\bibfield  {journal} {\bibinfo
  {journal} {Phys. Rev. C}\ }\textbf {\bibinfo {volume} {61}},\ \bibinfo
  {pages} {051303} (\bibinfo {year} {2000})}\BibitemShut {NoStop}%
\bibitem [{\citenamefont {Naidon}\ and\ \citenamefont
  {Endo}(2017)}]{Naidon:2016dpf}%
  \BibitemOpen
  \bibfield  {author} {\bibinfo {author} {\bibfnamefont {P.}~\bibnamefont
  {Naidon}}\ and\ \bibinfo {author} {\bibfnamefont {S.}~\bibnamefont {Endo}},\
  }\href {\doibase 10.1088/1361-6633/aa50e8} {\bibfield  {journal} {\bibinfo
  {journal} {Rept. Prog. Phys.}\ }\textbf {\bibinfo {volume} {80}},\ \bibinfo
  {pages} {056001} (\bibinfo {year} {2017})},\ \Eprint
  {http://arxiv.org/abs/1610.09805} {arXiv:1610.09805 [quant-ph]} \BibitemShut
  {NoStop}%
\bibitem [{\citenamefont {Braaten}\ and\ \citenamefont
  {Kusunoki}(2004)}]{Braaten:2003he}%
  \BibitemOpen
  \bibfield  {author} {\bibinfo {author} {\bibfnamefont {E.}~\bibnamefont
  {Braaten}}\ and\ \bibinfo {author} {\bibfnamefont {M.}~\bibnamefont
  {Kusunoki}},\ }\href {\doibase 10.1103/PhysRevD.69.074005} {\bibfield
  {journal} {\bibinfo  {journal} {Phys. Rev. D}\ }\textbf {\bibinfo {volume}
  {69}},\ \bibinfo {pages} {074005} (\bibinfo {year} {2004})},\ \Eprint
  {http://arxiv.org/abs/hep-ph/0311147} {arXiv:hep-ph/0311147} \BibitemShut
  {NoStop}%
\bibitem [{\citenamefont {Hammer}\ and\ \citenamefont
  {Platter}(2010)}]{Hammer:2010kp}%
  \BibitemOpen
  \bibfield  {author} {\bibinfo {author} {\bibfnamefont {H.-W.}\ \bibnamefont
  {Hammer}}\ and\ \bibinfo {author} {\bibfnamefont {L.}~\bibnamefont
  {Platter}},\ }\href {\doibase 10.1146/annurev.nucl.012809.104439} {\bibfield
  {journal} {\bibinfo  {journal} {Ann. Rev. Nucl. Part. Sci.}\ }\textbf
  {\bibinfo {volume} {60}},\ \bibinfo {pages} {207} (\bibinfo {year} {2010})},\
  \Eprint {http://arxiv.org/abs/1001.1981} {arXiv:1001.1981 [nucl-th]}
  \BibitemShut {NoStop}%
\bibitem [{\citenamefont {Braaten}\ and\ \citenamefont
  {Hammer}(2006)}]{Braaten:2004rn}%
  \BibitemOpen
  \bibfield  {author} {\bibinfo {author} {\bibfnamefont {E.}~\bibnamefont
  {Braaten}}\ and\ \bibinfo {author} {\bibfnamefont {H.~W.}\ \bibnamefont
  {Hammer}},\ }\href {\doibase 10.1016/j.physrep.2006.03.001} {\bibfield
  {journal} {\bibinfo  {journal} {Phys. Rept.}\ }\textbf {\bibinfo {volume}
  {428}},\ \bibinfo {pages} {259} (\bibinfo {year} {2006})},\ \Eprint
  {http://arxiv.org/abs/cond-mat/0410417} {arXiv:cond-mat/0410417} \BibitemShut
  {NoStop}%
\bibitem [{\citenamefont {Frederico}\ \emph {et~al.}(1999)\citenamefont
  {Frederico}, \citenamefont {Tomio}, \citenamefont {Delfino},\ and\
  \citenamefont {Amorim}}]{PhysRevA.60.R9}%
  \BibitemOpen
  \bibfield  {author} {\bibinfo {author} {\bibfnamefont {T.}~\bibnamefont
  {Frederico}}, \bibinfo {author} {\bibfnamefont {L.}~\bibnamefont {Tomio}},
  \bibinfo {author} {\bibfnamefont {A.}~\bibnamefont {Delfino}}, \ and\
  \bibinfo {author} {\bibfnamefont {A.~E.~A.}\ \bibnamefont {Amorim}},\ }\href
  {\doibase 10.1103/PhysRevA.60.R9} {\bibfield  {journal} {\bibinfo  {journal}
  {Phys. Rev. A}\ }\textbf {\bibinfo {volume} {60}},\ \bibinfo {pages} {R9}
  (\bibinfo {year} {1999})}\BibitemShut {NoStop}%
\bibitem [{\citenamefont {Dawid}\ \emph {et~al.}(2024)\citenamefont {Dawid},
  \citenamefont {Islam}, \citenamefont {Briceno},\ and\ \citenamefont
  {Jackura}}]{Dawid:2023kxu}%
  \BibitemOpen
  \bibfield  {author} {\bibinfo {author} {\bibfnamefont {S.~M.}\ \bibnamefont
  {Dawid}}, \bibinfo {author} {\bibfnamefont {M.~H.~E.}\ \bibnamefont {Islam}},
  \bibinfo {author} {\bibfnamefont {R.~A.}\ \bibnamefont {Briceno}}, \ and\
  \bibinfo {author} {\bibfnamefont {A.~W.}\ \bibnamefont {Jackura}},\ }\href
  {\doibase 10.1103/PhysRevA.109.043325} {\bibfield  {journal} {\bibinfo
  {journal} {Phys. Rev. A}\ }\textbf {\bibinfo {volume} {109}},\ \bibinfo
  {pages} {043325} (\bibinfo {year} {2024})},\ \Eprint
  {http://arxiv.org/abs/2309.01732} {arXiv:2309.01732 [nucl-th]} \BibitemShut
  {NoStop}%
\bibitem [{\citenamefont {Canham}\ \emph {et~al.}(2009)\citenamefont {Canham},
  \citenamefont {Hammer},\ and\ \citenamefont {Springer}}]{Canham:2009zq}%
  \BibitemOpen
  \bibfield  {author} {\bibinfo {author} {\bibfnamefont {D.~L.}\ \bibnamefont
  {Canham}}, \bibinfo {author} {\bibfnamefont {H.~W.}\ \bibnamefont {Hammer}},
  \ and\ \bibinfo {author} {\bibfnamefont {R.~P.}\ \bibnamefont {Springer}},\
  }\href {\doibase 10.1103/PhysRevD.80.014009} {\bibfield  {journal} {\bibinfo
  {journal} {Phys. Rev. D}\ }\textbf {\bibinfo {volume} {80}},\ \bibinfo
  {pages} {014009} (\bibinfo {year} {2009})},\ \Eprint
  {http://arxiv.org/abs/0906.1263} {arXiv:0906.1263 [hep-ph]} \BibitemShut
  {NoStop}%
\bibitem [{\citenamefont {Wilbring}\ \emph {et~al.}(2017)\citenamefont
  {Wilbring}, \citenamefont {Hammer},\ and\ \citenamefont
  {Mei\ss{}ner}}]{Wilbring:2017fwy}%
  \BibitemOpen
  \bibfield  {author} {\bibinfo {author} {\bibfnamefont {E.}~\bibnamefont
  {Wilbring}}, \bibinfo {author} {\bibfnamefont {H.~W.}\ \bibnamefont
  {Hammer}}, \ and\ \bibinfo {author} {\bibfnamefont {U.-G.}\ \bibnamefont
  {Mei\ss{}ner}},\ }\href@noop {} {\  (\bibinfo {year} {2017})},\ \Eprint
  {http://arxiv.org/abs/1705.06176} {arXiv:1705.06176 [hep-ph]} \BibitemShut
  {NoStop}%
\bibitem [{\citenamefont {Valderrama}(2019)}]{Valderrama:2018azi}%
  \BibitemOpen
  \bibfield  {author} {\bibinfo {author} {\bibfnamefont {M.~P.}\ \bibnamefont
  {Valderrama}},\ }\href {\doibase 10.1103/PhysRevD.99.034010} {\bibfield
  {journal} {\bibinfo  {journal} {Phys. Rev. D}\ }\textbf {\bibinfo {volume}
  {99}},\ \bibinfo {pages} {034010} (\bibinfo {year} {2019})},\ \Eprint
  {http://arxiv.org/abs/1811.10173} {arXiv:1811.10173 [hep-ph]} \BibitemShut
  {NoStop}%
\bibitem [{\citenamefont {Martinez~Torres}\ \emph {et~al.}(2020)\citenamefont
  {Martinez~Torres}, \citenamefont {Khemchandani}, \citenamefont {Roca},\ and\
  \citenamefont {Oset}}]{MartinezTorres:2020hus}%
  \BibitemOpen
  \bibfield  {author} {\bibinfo {author} {\bibfnamefont {A.}~\bibnamefont
  {Martinez~Torres}}, \bibinfo {author} {\bibfnamefont {K.~P.}\ \bibnamefont
  {Khemchandani}}, \bibinfo {author} {\bibfnamefont {L.}~\bibnamefont {Roca}},
  \ and\ \bibinfo {author} {\bibfnamefont {E.}~\bibnamefont {Oset}},\ }\href
  {\doibase 10.1007/s00601-020-01568-y} {\bibfield  {journal} {\bibinfo
  {journal} {Few Body Syst.}\ }\textbf {\bibinfo {volume} {61}},\ \bibinfo
  {pages} {35} (\bibinfo {year} {2020})},\ \Eprint
  {http://arxiv.org/abs/2005.14357} {arXiv:2005.14357 [nucl-th]} \BibitemShut
  {NoStop}%
\bibitem [{\citenamefont {Choi}\ \emph {et~al.}(2003)\citenamefont {Choi} \emph
  {et~al.}}]{Belle:2003nnu}%
  \BibitemOpen
  \bibfield  {author} {\bibinfo {author} {\bibfnamefont {S.~K.}\ \bibnamefont
  {Choi}} \emph {et~al.} (\bibinfo {collaboration} {Belle}),\ }\href {\doibase
  10.1103/PhysRevLett.91.262001} {\bibfield  {journal} {\bibinfo  {journal}
  {Phys. Rev. Lett.}\ }\textbf {\bibinfo {volume} {91}},\ \bibinfo {pages}
  {262001} (\bibinfo {year} {2003})},\ \Eprint
  {http://arxiv.org/abs/hep-ex/0309032} {arXiv:hep-ex/0309032} \BibitemShut
  {NoStop}%
\bibitem [{\citenamefont {Aaij}\ \emph
  {et~al.}(2022{\natexlab{a}})\citenamefont {Aaij} \emph
  {et~al.}}]{LHCb:2021vvq}%
  \BibitemOpen
  \bibfield  {author} {\bibinfo {author} {\bibfnamefont {R.}~\bibnamefont
  {Aaij}} \emph {et~al.} (\bibinfo {collaboration} {LHCb}),\ }\href {\doibase
  10.1038/s41567-022-01614-y} {\bibfield  {journal} {\bibinfo  {journal}
  {Nature Phys.}\ }\textbf {\bibinfo {volume} {18}},\ \bibinfo {pages} {751}
  (\bibinfo {year} {2022}{\natexlab{a}})},\ \Eprint
  {http://arxiv.org/abs/2109.01038} {arXiv:2109.01038 [hep-ex]} \BibitemShut
  {NoStop}%
\bibitem [{\citenamefont {Aaij}\ \emph
  {et~al.}(2022{\natexlab{b}})\citenamefont {Aaij} \emph
  {et~al.}}]{LHCb:2021auc}%
  \BibitemOpen
  \bibfield  {author} {\bibinfo {author} {\bibfnamefont {R.}~\bibnamefont
  {Aaij}} \emph {et~al.} (\bibinfo {collaboration} {LHCb}),\ }\href {\doibase
  10.1038/s41467-022-30206-w} {\bibfield  {journal} {\bibinfo  {journal}
  {Nature Commun.}\ }\textbf {\bibinfo {volume} {13}},\ \bibinfo {pages} {3351}
  (\bibinfo {year} {2022}{\natexlab{b}})},\ \Eprint
  {http://arxiv.org/abs/2109.01056} {arXiv:2109.01056 [hep-ex]} \BibitemShut
  {NoStop}%
\bibitem [{\citenamefont {Chen}\ \emph {et~al.}(2022)\citenamefont {Chen},
  \citenamefont {Chen}, \citenamefont {Liu}, \citenamefont {Liu},\ and\
  \citenamefont {Zhu}}]{Chen:2022asf}%
  \BibitemOpen
  \bibfield  {author} {\bibinfo {author} {\bibfnamefont {H.-X.}\ \bibnamefont
  {Chen}}, \bibinfo {author} {\bibfnamefont {W.}~\bibnamefont {Chen}}, \bibinfo
  {author} {\bibfnamefont {X.}~\bibnamefont {Liu}}, \bibinfo {author}
  {\bibfnamefont {Y.-R.}\ \bibnamefont {Liu}}, \ and\ \bibinfo {author}
  {\bibfnamefont {S.-L.}\ \bibnamefont {Zhu}},\ }\href@noop {} {\  (\bibinfo
  {year} {2022})},\ \Eprint {http://arxiv.org/abs/2204.02649} {arXiv:2204.02649
  [hep-ph]} \BibitemShut {NoStop}%
\bibitem [{\citenamefont {Wu}\ \emph {et~al.}(2022)\citenamefont {Wu},
  \citenamefont {Pan}, \citenamefont {Liu}, \citenamefont {Luo}, \citenamefont
  {Geng},\ and\ \citenamefont {Liu}}]{Wu:2021kbu}%
  \BibitemOpen
  \bibfield  {author} {\bibinfo {author} {\bibfnamefont {T.-W.}\ \bibnamefont
  {Wu}}, \bibinfo {author} {\bibfnamefont {Y.-W.}\ \bibnamefont {Pan}},
  \bibinfo {author} {\bibfnamefont {M.-Z.}\ \bibnamefont {Liu}}, \bibinfo
  {author} {\bibfnamefont {S.-Q.}\ \bibnamefont {Luo}}, \bibinfo {author}
  {\bibfnamefont {L.-S.}\ \bibnamefont {Geng}}, \ and\ \bibinfo {author}
  {\bibfnamefont {X.}~\bibnamefont {Liu}},\ }\href {\doibase
  10.1103/PhysRevD.105.L031505} {\bibfield  {journal} {\bibinfo  {journal}
  {Phys. Rev. D}\ }\textbf {\bibinfo {volume} {105}},\ \bibinfo {pages}
  {L031505} (\bibinfo {year} {2022})},\ \Eprint
  {http://arxiv.org/abs/2108.00923} {arXiv:2108.00923 [hep-ph]} \BibitemShut
  {NoStop}%
\bibitem [{\citenamefont {Pan}\ \emph {et~al.}(2022)\citenamefont {Pan},
  \citenamefont {Wu}, \citenamefont {Liu},\ and\ \citenamefont
  {Geng}}]{Pan:2022whr}%
  \BibitemOpen
  \bibfield  {author} {\bibinfo {author} {\bibfnamefont {Y.-W.}\ \bibnamefont
  {Pan}}, \bibinfo {author} {\bibfnamefont {T.-W.}\ \bibnamefont {Wu}},
  \bibinfo {author} {\bibfnamefont {M.-Z.}\ \bibnamefont {Liu}}, \ and\
  \bibinfo {author} {\bibfnamefont {L.-S.}\ \bibnamefont {Geng}},\ }\href
  {\doibase 10.1140/epjc/s10052-022-10881-1} {\bibfield  {journal} {\bibinfo
  {journal} {Eur. Phys. J. C}\ }\textbf {\bibinfo {volume} {82}},\ \bibinfo
  {pages} {908} (\bibinfo {year} {2022})},\ \Eprint
  {http://arxiv.org/abs/2208.05385} {arXiv:2208.05385 [hep-ph]} \BibitemShut
  {NoStop}%
\bibitem [{\citenamefont {Luo}\ \emph {et~al.}(2022)\citenamefont {Luo},
  \citenamefont {Wu}, \citenamefont {Liu}, \citenamefont {Geng},\ and\
  \citenamefont {Liu}}]{Luo:2021ggs}%
  \BibitemOpen
  \bibfield  {author} {\bibinfo {author} {\bibfnamefont {S.-Q.}\ \bibnamefont
  {Luo}}, \bibinfo {author} {\bibfnamefont {T.-W.}\ \bibnamefont {Wu}},
  \bibinfo {author} {\bibfnamefont {M.-Z.}\ \bibnamefont {Liu}}, \bibinfo
  {author} {\bibfnamefont {L.-S.}\ \bibnamefont {Geng}}, \ and\ \bibinfo
  {author} {\bibfnamefont {X.}~\bibnamefont {Liu}},\ }\href {\doibase
  10.1103/PhysRevD.105.074033} {\bibfield  {journal} {\bibinfo  {journal}
  {Phys. Rev. D}\ }\textbf {\bibinfo {volume} {105}},\ \bibinfo {pages}
  {074033} (\bibinfo {year} {2022})},\ \Eprint
  {http://arxiv.org/abs/2111.15079} {arXiv:2111.15079 [hep-ph]} \BibitemShut
  {NoStop}%
\bibitem [{\citenamefont {Bayar}\ \emph {et~al.}(2023)\citenamefont {Bayar},
  \citenamefont {Martinez~Torres}, \citenamefont {Khemchandani}, \citenamefont
  {Molina},\ and\ \citenamefont {Oset}}]{Bayar:2022bnc}%
  \BibitemOpen
  \bibfield  {author} {\bibinfo {author} {\bibfnamefont {M.}~\bibnamefont
  {Bayar}}, \bibinfo {author} {\bibfnamefont {A.}~\bibnamefont
  {Martinez~Torres}}, \bibinfo {author} {\bibfnamefont {K.~P.}\ \bibnamefont
  {Khemchandani}}, \bibinfo {author} {\bibfnamefont {R.}~\bibnamefont
  {Molina}}, \ and\ \bibinfo {author} {\bibfnamefont {E.}~\bibnamefont
  {Oset}},\ }\href {\doibase 10.1140/epjc/s10052-023-11207-5} {\bibfield
  {journal} {\bibinfo  {journal} {Eur. Phys. J. C}\ }\textbf {\bibinfo {volume}
  {83}},\ \bibinfo {pages} {46} (\bibinfo {year} {2023})},\ \Eprint
  {http://arxiv.org/abs/2211.09294} {arXiv:2211.09294 [hep-ph]} \BibitemShut
  {NoStop}%
\bibitem [{\citenamefont {Neubert}(1994)}]{Neubert:1993mb}%
  \BibitemOpen
  \bibfield  {author} {\bibinfo {author} {\bibfnamefont {M.}~\bibnamefont
  {Neubert}},\ }\href {\doibase 10.1016/0370-1573(94)90091-4} {\bibfield
  {journal} {\bibinfo  {journal} {Phys. Rept.}\ }\textbf {\bibinfo {volume}
  {245}},\ \bibinfo {pages} {259} (\bibinfo {year} {1994})},\ \Eprint
  {http://arxiv.org/abs/hep-ph/9306320} {arXiv:hep-ph/9306320} \BibitemShut
  {NoStop}%
\bibitem [{\citenamefont {Albaladejo}(2022)}]{Albaladejo:2021vln}%
  \BibitemOpen
  \bibfield  {author} {\bibinfo {author} {\bibfnamefont {M.}~\bibnamefont
  {Albaladejo}},\ }\href {\doibase 10.1016/j.physletb.2022.137052} {\bibfield
  {journal} {\bibinfo  {journal} {Phys. Lett. B}\ }\textbf {\bibinfo {volume}
  {829}},\ \bibinfo {pages} {137052} (\bibinfo {year} {2022})},\ \Eprint
  {http://arxiv.org/abs/2110.02944} {arXiv:2110.02944 [hep-ph]} \BibitemShut
  {NoStop}%
\bibitem [{\citenamefont {Liu}\ \emph {et~al.}(2019)\citenamefont {Liu},
  \citenamefont {Wu}, \citenamefont {Pavon~Valderrama}, \citenamefont {Xie},\
  and\ \citenamefont {Geng}}]{Liu:2019stu}%
  \BibitemOpen
  \bibfield  {author} {\bibinfo {author} {\bibfnamefont {M.-Z.}\ \bibnamefont
  {Liu}}, \bibinfo {author} {\bibfnamefont {T.-W.}\ \bibnamefont {Wu}},
  \bibinfo {author} {\bibfnamefont {M.}~\bibnamefont {Pavon~Valderrama}},
  \bibinfo {author} {\bibfnamefont {J.-J.}\ \bibnamefont {Xie}}, \ and\
  \bibinfo {author} {\bibfnamefont {L.-S.}\ \bibnamefont {Geng}},\ }\href
  {\doibase 10.1103/PhysRevD.99.094018} {\bibfield  {journal} {\bibinfo
  {journal} {Phys. Rev. D}\ }\textbf {\bibinfo {volume} {99}},\ \bibinfo
  {pages} {094018} (\bibinfo {year} {2019})},\ \Eprint
  {http://arxiv.org/abs/1902.03044} {arXiv:1902.03044 [hep-ph]} \BibitemShut
  {NoStop}%
\bibitem [{\citenamefont {Dai}\ \emph {et~al.}(2022)\citenamefont {Dai},
  \citenamefont {Molina},\ and\ \citenamefont {Oset}}]{Dai:2021vgf}%
  \BibitemOpen
  \bibfield  {author} {\bibinfo {author} {\bibfnamefont {L.~R.}\ \bibnamefont
  {Dai}}, \bibinfo {author} {\bibfnamefont {R.}~\bibnamefont {Molina}}, \ and\
  \bibinfo {author} {\bibfnamefont {E.}~\bibnamefont {Oset}},\ }\href {\doibase
  10.1103/PhysRevD.105.016029} {\bibfield  {journal} {\bibinfo  {journal}
  {Phys. Rev. D}\ }\textbf {\bibinfo {volume} {105}},\ \bibinfo {pages}
  {016029} (\bibinfo {year} {2022})},\ \bibinfo {note} {[Erratum: Phys.Rev.D
  106, 099902 (2022)]},\ \Eprint {http://arxiv.org/abs/2110.15270}
  {arXiv:2110.15270 [hep-ph]} \BibitemShut {NoStop}%
\bibitem [{\citenamefont {Du}\ \emph {et~al.}(2022)\citenamefont {Du},
  \citenamefont {Baru}, \citenamefont {Dong}, \citenamefont {Filin},
  \citenamefont {Guo}, \citenamefont {Hanhart}, \citenamefont {Nefediev},
  \citenamefont {Nieves},\ and\ \citenamefont {Wang}}]{Du:2021zzh}%
  \BibitemOpen
  \bibfield  {author} {\bibinfo {author} {\bibfnamefont {M.-L.}\ \bibnamefont
  {Du}}, \bibinfo {author} {\bibfnamefont {V.}~\bibnamefont {Baru}}, \bibinfo
  {author} {\bibfnamefont {X.-K.}\ \bibnamefont {Dong}}, \bibinfo {author}
  {\bibfnamefont {A.}~\bibnamefont {Filin}}, \bibinfo {author} {\bibfnamefont
  {F.-K.}\ \bibnamefont {Guo}}, \bibinfo {author} {\bibfnamefont
  {C.}~\bibnamefont {Hanhart}}, \bibinfo {author} {\bibfnamefont
  {A.}~\bibnamefont {Nefediev}}, \bibinfo {author} {\bibfnamefont
  {J.}~\bibnamefont {Nieves}}, \ and\ \bibinfo {author} {\bibfnamefont
  {Q.}~\bibnamefont {Wang}},\ }\href {\doibase 10.1103/PhysRevD.105.014024}
  {\bibfield  {journal} {\bibinfo  {journal} {Phys. Rev. D}\ }\textbf {\bibinfo
  {volume} {105}},\ \bibinfo {pages} {014024} (\bibinfo {year} {2022})},\
  \Eprint {http://arxiv.org/abs/2110.13765} {arXiv:2110.13765 [hep-ph]}
  \BibitemShut {NoStop}%
\bibitem [{\citenamefont {Dong}\ \emph {et~al.}(2021)\citenamefont {Dong},
  \citenamefont {Guo},\ and\ \citenamefont {Zou}}]{Dong:2021bvy}%
  \BibitemOpen
  \bibfield  {author} {\bibinfo {author} {\bibfnamefont {X.-K.}\ \bibnamefont
  {Dong}}, \bibinfo {author} {\bibfnamefont {F.-K.}\ \bibnamefont {Guo}}, \
  and\ \bibinfo {author} {\bibfnamefont {B.-S.}\ \bibnamefont {Zou}},\ }\href
  {\doibase 10.1088/1572-9494/ac27a2} {\bibfield  {journal} {\bibinfo
  {journal} {Commun. Theor. Phys.}\ }\textbf {\bibinfo {volume} {73}},\
  \bibinfo {pages} {125201} (\bibinfo {year} {2021})},\ \Eprint
  {http://arxiv.org/abs/2108.02673} {arXiv:2108.02673 [hep-ph]} \BibitemShut
  {NoStop}%
\bibitem [{\citenamefont {Qiu}\ \emph {et~al.}(2023)\citenamefont {Qiu},
  \citenamefont {Gong},\ and\ \citenamefont {Zhao}}]{Qiu:2023uno}%
  \BibitemOpen
  \bibfield  {author} {\bibinfo {author} {\bibfnamefont {L.}~\bibnamefont
  {Qiu}}, \bibinfo {author} {\bibfnamefont {C.}~\bibnamefont {Gong}}, \ and\
  \bibinfo {author} {\bibfnamefont {Q.}~\bibnamefont {Zhao}},\ }\href@noop {}
  {\  (\bibinfo {year} {2023})},\ \Eprint {http://arxiv.org/abs/2311.10067}
  {arXiv:2311.10067 [hep-ph]} \BibitemShut {NoStop}%
\bibitem [{\citenamefont {Peng}\ \emph {et~al.}(2023)\citenamefont {Peng},
  \citenamefont {Yan},\ and\ \citenamefont {Pavon~Valderrama}}]{Peng:2023lfw}%
  \BibitemOpen
  \bibfield  {author} {\bibinfo {author} {\bibfnamefont {F.-Z.}\ \bibnamefont
  {Peng}}, \bibinfo {author} {\bibfnamefont {M.-J.}\ \bibnamefont {Yan}}, \
  and\ \bibinfo {author} {\bibfnamefont {M.}~\bibnamefont {Pavon~Valderrama}},\
  }\href {\doibase 10.1103/PhysRevD.108.114001} {\bibfield  {journal} {\bibinfo
   {journal} {Phys. Rev. D}\ }\textbf {\bibinfo {volume} {108}},\ \bibinfo
  {pages} {114001} (\bibinfo {year} {2023})},\ \Eprint
  {http://arxiv.org/abs/2304.13515} {arXiv:2304.13515 [hep-ph]} \BibitemShut
  {NoStop}%
\bibitem [{\citenamefont {Abreu}(2022)}]{Abreu:2022sra}%
  \BibitemOpen
  \bibfield  {author} {\bibinfo {author} {\bibfnamefont {L.~M.}\ \bibnamefont
  {Abreu}},\ }\href {\doibase 10.1016/j.nuclphysb.2022.115994} {\bibfield
  {journal} {\bibinfo  {journal} {Nucl. Phys. B}\ }\textbf {\bibinfo {volume}
  {985}},\ \bibinfo {pages} {115994} (\bibinfo {year} {2022})},\ \Eprint
  {http://arxiv.org/abs/2206.01166} {arXiv:2206.01166 [hep-ph]} \BibitemShut
  {NoStop}%
\bibitem [{\citenamefont {Ortega}\ \emph {et~al.}(2023)\citenamefont {Ortega},
  \citenamefont {Segovia}, \citenamefont {Entem},\ and\ \citenamefont
  {Fernandez}}]{Ortega:2022efc}%
  \BibitemOpen
  \bibfield  {author} {\bibinfo {author} {\bibfnamefont {P.~G.}\ \bibnamefont
  {Ortega}}, \bibinfo {author} {\bibfnamefont {J.}~\bibnamefont {Segovia}},
  \bibinfo {author} {\bibfnamefont {D.~R.}\ \bibnamefont {Entem}}, \ and\
  \bibinfo {author} {\bibfnamefont {F.}~\bibnamefont {Fernandez}},\ }\href
  {\doibase 10.1016/j.physletb.2023.137918} {\bibfield  {journal} {\bibinfo
  {journal} {Phys. Lett. B}\ }\textbf {\bibinfo {volume} {841}},\ \bibinfo
  {pages} {137918} (\bibinfo {year} {2023})},\ \bibinfo {note} {[Erratum:
  Phys.Lett.B 847, 138308 (2023)]},\ \Eprint {http://arxiv.org/abs/2211.06118}
  {arXiv:2211.06118 [hep-ph]} \BibitemShut {NoStop}%
\bibitem [{\citenamefont {Li}\ \emph {et~al.}(2013)\citenamefont {Li},
  \citenamefont {Sun}, \citenamefont {Liu},\ and\ \citenamefont
  {Zhu}}]{Li:2012ss}%
  \BibitemOpen
  \bibfield  {author} {\bibinfo {author} {\bibfnamefont {N.}~\bibnamefont
  {Li}}, \bibinfo {author} {\bibfnamefont {Z.-F.}\ \bibnamefont {Sun}},
  \bibinfo {author} {\bibfnamefont {X.}~\bibnamefont {Liu}}, \ and\ \bibinfo
  {author} {\bibfnamefont {S.-L.}\ \bibnamefont {Zhu}},\ }\href {\doibase
  10.1103/PhysRevD.88.114008} {\bibfield  {journal} {\bibinfo  {journal} {Phys.
  Rev. D}\ }\textbf {\bibinfo {volume} {88}},\ \bibinfo {pages} {114008}
  (\bibinfo {year} {2013})},\ \Eprint {http://arxiv.org/abs/1211.5007}
  {arXiv:1211.5007 [hep-ph]} \BibitemShut {NoStop}%
\bibitem [{\citenamefont {Molina}\ \emph {et~al.}(2010)\citenamefont {Molina},
  \citenamefont {Branz},\ and\ \citenamefont {Oset}}]{Molina:2010tx}%
  \BibitemOpen
  \bibfield  {author} {\bibinfo {author} {\bibfnamefont {R.}~\bibnamefont
  {Molina}}, \bibinfo {author} {\bibfnamefont {T.}~\bibnamefont {Branz}}, \
  and\ \bibinfo {author} {\bibfnamefont {E.}~\bibnamefont {Oset}},\ }\href
  {\doibase 10.1103/PhysRevD.82.014010} {\bibfield  {journal} {\bibinfo
  {journal} {Phys. Rev. D}\ }\textbf {\bibinfo {volume} {82}},\ \bibinfo
  {pages} {014010} (\bibinfo {year} {2010})},\ \Eprint
  {http://arxiv.org/abs/1005.0335} {arXiv:1005.0335 [hep-ph]} \BibitemShut
  {NoStop}%
\bibitem [{\citenamefont {Jia}\ \emph {et~al.}(2023)\citenamefont {Jia},
  \citenamefont {Yan}, \citenamefont {Zhang}, \citenamefont {Shi},
  \citenamefont {Li},\ and\ \citenamefont {Guo}}]{Jia:2022qwr}%
  \BibitemOpen
  \bibfield  {author} {\bibinfo {author} {\bibfnamefont {Z.-S.}\ \bibnamefont
  {Jia}}, \bibinfo {author} {\bibfnamefont {M.-J.}\ \bibnamefont {Yan}},
  \bibinfo {author} {\bibfnamefont {Z.-H.}\ \bibnamefont {Zhang}}, \bibinfo
  {author} {\bibfnamefont {P.-P.}\ \bibnamefont {Shi}}, \bibinfo {author}
  {\bibfnamefont {G.}~\bibnamefont {Li}}, \ and\ \bibinfo {author}
  {\bibfnamefont {F.-K.}\ \bibnamefont {Guo}},\ }\href {\doibase
  10.1103/PhysRevD.107.074029} {\bibfield  {journal} {\bibinfo  {journal}
  {Phys. Rev. D}\ }\textbf {\bibinfo {volume} {107}},\ \bibinfo {pages}
  {074029} (\bibinfo {year} {2023})},\ \Eprint
  {http://arxiv.org/abs/2211.02479} {arXiv:2211.02479 [hep-ph]} \BibitemShut
  {NoStop}%
\bibitem [{\citenamefont {Whyte}\ \emph {et~al.}(2024)\citenamefont {Whyte},
  \citenamefont {Wilson},\ and\ \citenamefont {Thomas}}]{Whyte:2024ihh}%
  \BibitemOpen
  \bibfield  {author} {\bibinfo {author} {\bibfnamefont {T.}~\bibnamefont
  {Whyte}}, \bibinfo {author} {\bibfnamefont {D.~J.}\ \bibnamefont {Wilson}}, \
  and\ \bibinfo {author} {\bibfnamefont {C.~E.}\ \bibnamefont {Thomas}},\
  }\href@noop {} {\  (\bibinfo {year} {2024})},\ \Eprint
  {http://arxiv.org/abs/2405.15741} {arXiv:2405.15741 [hep-lat]} \BibitemShut
  {NoStop}%
\bibitem [{\citenamefont {Nieves}\ and\ \citenamefont
  {Ruiz~Arriola}(2000)}]{Nieves:1999bx}%
  \BibitemOpen
  \bibfield  {author} {\bibinfo {author} {\bibfnamefont {J.}~\bibnamefont
  {Nieves}}\ and\ \bibinfo {author} {\bibfnamefont {E.}~\bibnamefont
  {Ruiz~Arriola}},\ }\href {\doibase 10.1016/S0375-9474(00)00321-3} {\bibfield
  {journal} {\bibinfo  {journal} {Nucl. Phys. A}\ }\textbf {\bibinfo {volume}
  {679}},\ \bibinfo {pages} {57} (\bibinfo {year} {2000})},\ \Eprint
  {http://arxiv.org/abs/hep-ph/9907469} {arXiv:hep-ph/9907469} \BibitemShut
  {NoStop}%
\bibitem [{\citenamefont {Oller}\ \emph {et~al.}(1999)\citenamefont {Oller},
  \citenamefont {Oset},\ and\ \citenamefont {Pelaez}}]{Oller:1998hw}%
  \BibitemOpen
  \bibfield  {author} {\bibinfo {author} {\bibfnamefont {J.~A.}\ \bibnamefont
  {Oller}}, \bibinfo {author} {\bibfnamefont {E.}~\bibnamefont {Oset}}, \ and\
  \bibinfo {author} {\bibfnamefont {J.~R.}\ \bibnamefont {Pelaez}},\ }\href
  {\doibase 10.1103/PhysRevD.59.074001} {\bibfield  {journal} {\bibinfo
  {journal} {Phys. Rev. D}\ }\textbf {\bibinfo {volume} {59}},\ \bibinfo
  {pages} {074001} (\bibinfo {year} {1999})},\ \bibinfo {note} {[Erratum:
  Phys.Rev.D 60, 099906 (1999), Erratum: Phys.Rev.D 75, 099903 (2007)]},\
  \Eprint {http://arxiv.org/abs/hep-ph/9804209} {arXiv:hep-ph/9804209}
  \BibitemShut {NoStop}%
\bibitem [{\citenamefont {Montesinos}\ \emph {et~al.}(2023)\citenamefont
  {Montesinos}, \citenamefont {Albaladejo}, \citenamefont {Nieves},\ and\
  \citenamefont {Tolos}}]{Montesinos:2023qbx}%
  \BibitemOpen
  \bibfield  {author} {\bibinfo {author} {\bibfnamefont {V.}~\bibnamefont
  {Montesinos}}, \bibinfo {author} {\bibfnamefont {M.}~\bibnamefont
  {Albaladejo}}, \bibinfo {author} {\bibfnamefont {J.}~\bibnamefont {Nieves}},
  \ and\ \bibinfo {author} {\bibfnamefont {L.}~\bibnamefont {Tolos}},\ }\href
  {\doibase 10.1103/PhysRevC.108.035205} {\bibfield  {journal} {\bibinfo
  {journal} {Phys. Rev. C}\ }\textbf {\bibinfo {volume} {108}},\ \bibinfo
  {pages} {035205} (\bibinfo {year} {2023})},\ \Eprint
  {http://arxiv.org/abs/2306.17673} {arXiv:2306.17673 [hep-ph]} \BibitemShut
  {NoStop}%
\bibitem [{\citenamefont {Hansen}\ and\ \citenamefont
  {Sharpe}(2015)}]{Hansen:2015zga}%
  \BibitemOpen
  \bibfield  {author} {\bibinfo {author} {\bibfnamefont {M.~T.}\ \bibnamefont
  {Hansen}}\ and\ \bibinfo {author} {\bibfnamefont {S.~R.}\ \bibnamefont
  {Sharpe}},\ }\href {\doibase 10.1103/PhysRevD.92.114509} {\bibfield
  {journal} {\bibinfo  {journal} {Phys. Rev. D}\ }\textbf {\bibinfo {volume}
  {92}},\ \bibinfo {pages} {114509} (\bibinfo {year} {2015})},\ \Eprint
  {http://arxiv.org/abs/1504.04248} {arXiv:1504.04248 [hep-lat]} \BibitemShut
  {NoStop}%
\bibitem [{\citenamefont {Jackura}\ \emph {et~al.}(2021)\citenamefont
  {Jackura}, \citenamefont {Brice\~no}, \citenamefont {Dawid}, \citenamefont
  {Islam},\ and\ \citenamefont {McCarty}}]{Jackura:2020bsk}%
  \BibitemOpen
  \bibfield  {author} {\bibinfo {author} {\bibfnamefont {A.~W.}\ \bibnamefont
  {Jackura}}, \bibinfo {author} {\bibfnamefont {R.~A.}\ \bibnamefont
  {Brice\~no}}, \bibinfo {author} {\bibfnamefont {S.~M.}\ \bibnamefont
  {Dawid}}, \bibinfo {author} {\bibfnamefont {M.~H.~E.}\ \bibnamefont {Islam}},
  \ and\ \bibinfo {author} {\bibfnamefont {C.}~\bibnamefont {McCarty}},\ }\href
  {\doibase 10.1103/PhysRevD.104.014507} {\bibfield  {journal} {\bibinfo
  {journal} {Phys. Rev. D}\ }\textbf {\bibinfo {volume} {104}},\ \bibinfo
  {pages} {014507} (\bibinfo {year} {2021})},\ \Eprint
  {http://arxiv.org/abs/2010.09820} {arXiv:2010.09820 [hep-lat]} \BibitemShut
  {NoStop}%
\bibitem [{\citenamefont {Dawid}\ \emph {et~al.}(2023)\citenamefont {Dawid},
  \citenamefont {Islam},\ and\ \citenamefont {Brice\~no}}]{Dawid:2023jrj}%
  \BibitemOpen
  \bibfield  {author} {\bibinfo {author} {\bibfnamefont {S.~M.}\ \bibnamefont
  {Dawid}}, \bibinfo {author} {\bibfnamefont {M.~H.~E.}\ \bibnamefont {Islam}},
  \ and\ \bibinfo {author} {\bibfnamefont {R.~A.}\ \bibnamefont {Brice\~no}},\
  }\href {\doibase 10.1103/PhysRevD.108.034016} {\bibfield  {journal} {\bibinfo
   {journal} {Phys. Rev. D}\ }\textbf {\bibinfo {volume} {108}},\ \bibinfo
  {pages} {034016} (\bibinfo {year} {2023})},\ \Eprint
  {http://arxiv.org/abs/2303.04394} {arXiv:2303.04394 [nucl-th]} \BibitemShut
  {NoStop}%
\bibitem [{\citenamefont {Martinez~Torres}\ \emph {et~al.}(2008)\citenamefont
  {Martinez~Torres}, \citenamefont {Khemchandani}, \citenamefont {Geng},
  \citenamefont {Napsuciale},\ and\ \citenamefont
  {Oset}}]{MartinezTorres:2008gy}%
  \BibitemOpen
  \bibfield  {author} {\bibinfo {author} {\bibfnamefont {A.}~\bibnamefont
  {Martinez~Torres}}, \bibinfo {author} {\bibfnamefont {K.~P.}\ \bibnamefont
  {Khemchandani}}, \bibinfo {author} {\bibfnamefont {L.~S.}\ \bibnamefont
  {Geng}}, \bibinfo {author} {\bibfnamefont {M.}~\bibnamefont {Napsuciale}}, \
  and\ \bibinfo {author} {\bibfnamefont {E.}~\bibnamefont {Oset}},\ }\href
  {\doibase 10.1103/PhysRevD.78.074031} {\bibfield  {journal} {\bibinfo
  {journal} {Phys. Rev. D}\ }\textbf {\bibinfo {volume} {78}},\ \bibinfo
  {pages} {074031} (\bibinfo {year} {2008})},\ \Eprint
  {http://arxiv.org/abs/0801.3635} {arXiv:0801.3635 [nucl-th]} \BibitemShut
  {NoStop}%
\bibitem [{\citenamefont {Khemchandani}\ \emph {et~al.}(2008)\citenamefont
  {Khemchandani}, \citenamefont {Martinez~Torres},\ and\ \citenamefont
  {Oset}}]{Khemchandani:2008rk}%
  \BibitemOpen
  \bibfield  {author} {\bibinfo {author} {\bibfnamefont {K.~P.}\ \bibnamefont
  {Khemchandani}}, \bibinfo {author} {\bibfnamefont {A.}~\bibnamefont
  {Martinez~Torres}}, \ and\ \bibinfo {author} {\bibfnamefont {E.}~\bibnamefont
  {Oset}},\ }\href {\doibase 10.1140/epja/i2008-10625-3} {\bibfield  {journal}
  {\bibinfo  {journal} {Eur. Phys. J. A}\ }\textbf {\bibinfo {volume} {37}},\
  \bibinfo {pages} {233} (\bibinfo {year} {2008})},\ \Eprint
  {http://arxiv.org/abs/0804.4670} {arXiv:0804.4670 [nucl-th]} \BibitemShut
  {NoStop}%
\bibitem [{\citenamefont {Jackura}\ \emph {et~al.}(2019)\citenamefont
  {Jackura}, \citenamefont {Dawid}, \citenamefont {Fern\'andez-Ram\'\i{}rez},
  \citenamefont {Mathieu}, \citenamefont {Mikhasenko}, \citenamefont {Pilloni},
  \citenamefont {Sharpe},\ and\ \citenamefont {Szczepaniak}}]{Jackura:2019bmu}%
  \BibitemOpen
  \bibfield  {author} {\bibinfo {author} {\bibfnamefont {A.~W.}\ \bibnamefont
  {Jackura}}, \bibinfo {author} {\bibfnamefont {S.~M.}\ \bibnamefont {Dawid}},
  \bibinfo {author} {\bibfnamefont {C.}~\bibnamefont
  {Fern\'andez-Ram\'\i{}rez}}, \bibinfo {author} {\bibfnamefont
  {V.}~\bibnamefont {Mathieu}}, \bibinfo {author} {\bibfnamefont
  {M.}~\bibnamefont {Mikhasenko}}, \bibinfo {author} {\bibfnamefont
  {A.}~\bibnamefont {Pilloni}}, \bibinfo {author} {\bibfnamefont {S.~R.}\
  \bibnamefont {Sharpe}}, \ and\ \bibinfo {author} {\bibfnamefont {A.~P.}\
  \bibnamefont {Szczepaniak}},\ }\href {\doibase 10.1103/PhysRevD.100.034508}
  {\bibfield  {journal} {\bibinfo  {journal} {Phys. Rev. D}\ }\textbf {\bibinfo
  {volume} {100}},\ \bibinfo {pages} {034508} (\bibinfo {year} {2019})},\
  \Eprint {http://arxiv.org/abs/1905.12007} {arXiv:1905.12007 [hep-ph]}
  \BibitemShut {NoStop}%
\bibitem [{\citenamefont {Giebink}(1985)}]{PhysRevC.32.502}%
  \BibitemOpen
  \bibfield  {author} {\bibinfo {author} {\bibfnamefont {D.~R.}\ \bibnamefont
  {Giebink}},\ }\href {\doibase 10.1103/PhysRevC.32.502} {\bibfield  {journal}
  {\bibinfo  {journal} {Phys. Rev. C}\ }\textbf {\bibinfo {volume} {32}},\
  \bibinfo {pages} {502} (\bibinfo {year} {1985})}\BibitemShut {NoStop}%
\bibitem [{\citenamefont {Jackura}\ and\ \citenamefont
  {Brice\~no}(2023)}]{Jackura:2023qtp}%
  \BibitemOpen
  \bibfield  {author} {\bibinfo {author} {\bibfnamefont {A.~W.}\ \bibnamefont
  {Jackura}}\ and\ \bibinfo {author} {\bibfnamefont {R.~A.}\ \bibnamefont
  {Brice\~no}},\ }\href@noop {} {\  (\bibinfo {year} {2023})},\ \Eprint
  {http://arxiv.org/abs/2312.00625} {arXiv:2312.00625 [hep-ph]} \BibitemShut
  {NoStop}%
\end{thebibliography}%

\end{document}